\documentclass[12pt]{article}

\usepackage[margin=2.0cm]{geometry}
\usepackage{amsmath}

\usepackage{multirow}
\usepackage{enumitem}
\usepackage{amssymb}
\usepackage{amsthm}
\usepackage{amsmath}
\usepackage{graphicx}
\usepackage{array} 
\usepackage{subcaption}
\usepackage{lipsum} 
\usepackage{setspace}
\usepackage[sort&compress,numbers]{natbib}
\usepackage[colorlinks=true, linkcolor=blue, citecolor=blue, urlcolor=blue]{hyperref} 

\usepackage{multirow}
\usepackage{booktabs}
\usepackage{tabularx}
\graphicspath{ {figures/} }
\usepackage{float}

\usepackage{tikz}
\usetikzlibrary{shapes.geometric, arrows.meta, positioning}
\usepackage{algorithm}
\usepackage{algorithmic}
\usepackage{pdflscape}
\usepackage{graphicx}

\usepackage{tikz}
\usetikzlibrary{shapes.geometric, arrows, positioning}

\tikzstyle{process} = [rectangle, minimum width=3cm, minimum height=1cm, text centered, draw=black]

\title{Sensitivity Analysis of Tactical Wireless Network Design Under Realistic Operational Constraints}
\author{
	Wissem Ahmed Zaid,
	Alain Hertz
	\\[3mm]
	\footnotesize Polytechnique Montr\'eal - Gerad, Montr\'eal, Canada\\[3mm]
}

\date{\today}

\begin{document}

\maketitle

\begin{abstract}
The design of tactical wireless networks reflects a complex interplay among structural constraints, technological choices, and underlying modeling assumptions. Although optimization-based approaches have been widely explored, the impact of configuration parameters on network topology quality and overall performance is still not fully understood.
This paper presents a comprehensive sensitivity analysis of tactical wireless network design under realistic operational constraints. It systematically investigates three categories of parameters: (i) structural topology rules, including master hub selection; (ii) technological factors such as antenna beamwidth; and (iii) modeling parameters embedded in the objective formulation.
Optimized topologies are produced using a Tabu Search metaheuristic, and statistical analyses based on the Friedman and Wilcoxon tests are performed to assess the significance of observed variations across different network sizes.
The findings reveal scale-dependent technological transitions and threshold effects in structural constraints. The analysis differentiates parameters that fundamentally reshape network topology from those that primarily influence performance magnitude. Together, these insights provide practical guidance for parameter tuning and topology configuration in mission-critical tactical wireless deployments.
\end{abstract}

\vspace{0.3cm}\noindent\emph{Keywords:} 
Tactical wireless network design, sensitivity analysis.

\section{Introduction}

Wireless communication has become an essential component of modern information systems, enabling connectivity across a wide range of critical applications. In scenarios where conventional telecommunication infrastructures are unavailable or fail such as during natural disasters or military operations, temporary tactical wireless networks are deployed to maintain essential communication. The central objective is to establish reliable links between key locations, ensuring efficient data exchange under strict structural and operational conditions.

This work addresses the design of tactical wireless networks arising from a concrete industrial application in the communications domain. The systems under consideration typically involve between 10 and 50 geographically distributed nodes. Every node is equipped with a radio interface connected to two multi-beam antennas. The radio operates with two distinct channels (one per antenna) and each channel can use two different transmission frequencies.
The network is structured as a rooted, directed tree, where all links are oriented outward from a designated root node. This root, referred to as the master hub, plays a central role by coordinating and managing communications across the network.
Two communication modes are supported between nodes. In point-to-point (PTP) mode, a dedicated link connects exactly two nodes, offering stable and high-quality transmission over longer distances. In contrast, point-to-multipoint (PMP) mode allows a single node to simultaneously serve multiple downstream nodes, which is particularly useful for distributing data efficiently over a broader area.
A key performance measure of the network is its bottleneck throughput, defined by the weakest link in terms of effective data rate. The objective is to maximize this bottleneck value, thereby improving the minimum transmission capacity across the entire network and ensuring reliable and efficient communication between all nodes.

This study provides an in-depth sensitivity analysis for tactical wireless network design. It examines how three broad categories of parameters affect the system: structural decisions related to the network configuration, technological features, and the modeling parameters involved in defining the optimization objective.
A more detailed description of the tactical wireless network design problem is provided in the next section, together with an overview of the main ingredients and factors that can influence the quality and performance of the resulting network configurations. Section \ref{sec:litterature} then presents a concise review of the existing literature on the design of such wireless networks, and a description of the statistical analysis methodology is given in Section \ref{sec:statistical_analysis}.
A sensitivity analysis is then performed in Sections \ref{sA:masterhub}--\ref{sec:sensitivity_objective}, where the corresponding results are presented, leading to the study’s conclusions in Section \ref{sec:conclusion}.

\section{Problem description}\label{sec:problem}
A detailed and comprehensive description of the problem can be found in \cite{WH1}. In the present context, however, we limit ourselves to a more general and less exhaustive account, retaining only those components and elements that are necessary and relevant for the purposes of the sensitivity analysis.
An instance of the problem is defined by a set $V$ of nodes, whose size typically ranges from 10 to 30, 
each node being associated with fixed spatial coordinates that determine its position in the network.
The network to be designed is required to have a tree topology. Once a tree connecting all nodes in $V$ has been constructed, one node is selected to serve as the master hub, and the tree is then oriented by directing all edges outward from this root node.
Each node is equipped with a radio connected to two multi-beam antennas. The radio has two channels, with each channel linked to a distinct antenna and capable of operating on one of two available frequencies. At the master hub, its immediate successors are partitioned into two groups, each group being assigned to one of the hub’s channels.
Every other node in the network has exactly one predecessor and may have several successors. One of its channels is reserved for communication with its predecessor, while the other channel is used to handle communication with its successors. If a node has only one successor, the corresponding link operates in point-to-point (PTP) mode; if it has two or more successors, the communication takes place in point-to-multipoint (PMP) mode.
The configuration of the antennas aims to minimize the number of active beams, since increasing the number of beams reduces the signal strength available to each connected link. Additionally, due to technological constraints, a radio interface operating on a single channel can communicate with at most ten other nodes. Consequently, the master hub can have a maximum degree of 20, while any other node is limited to a maximum degree of 11, comprising one link associated with its predecessor and up to ten links associated with its successors in the directed tree.

\begin{figure}[!htb]	\centering   \includegraphics[scale=0.5]{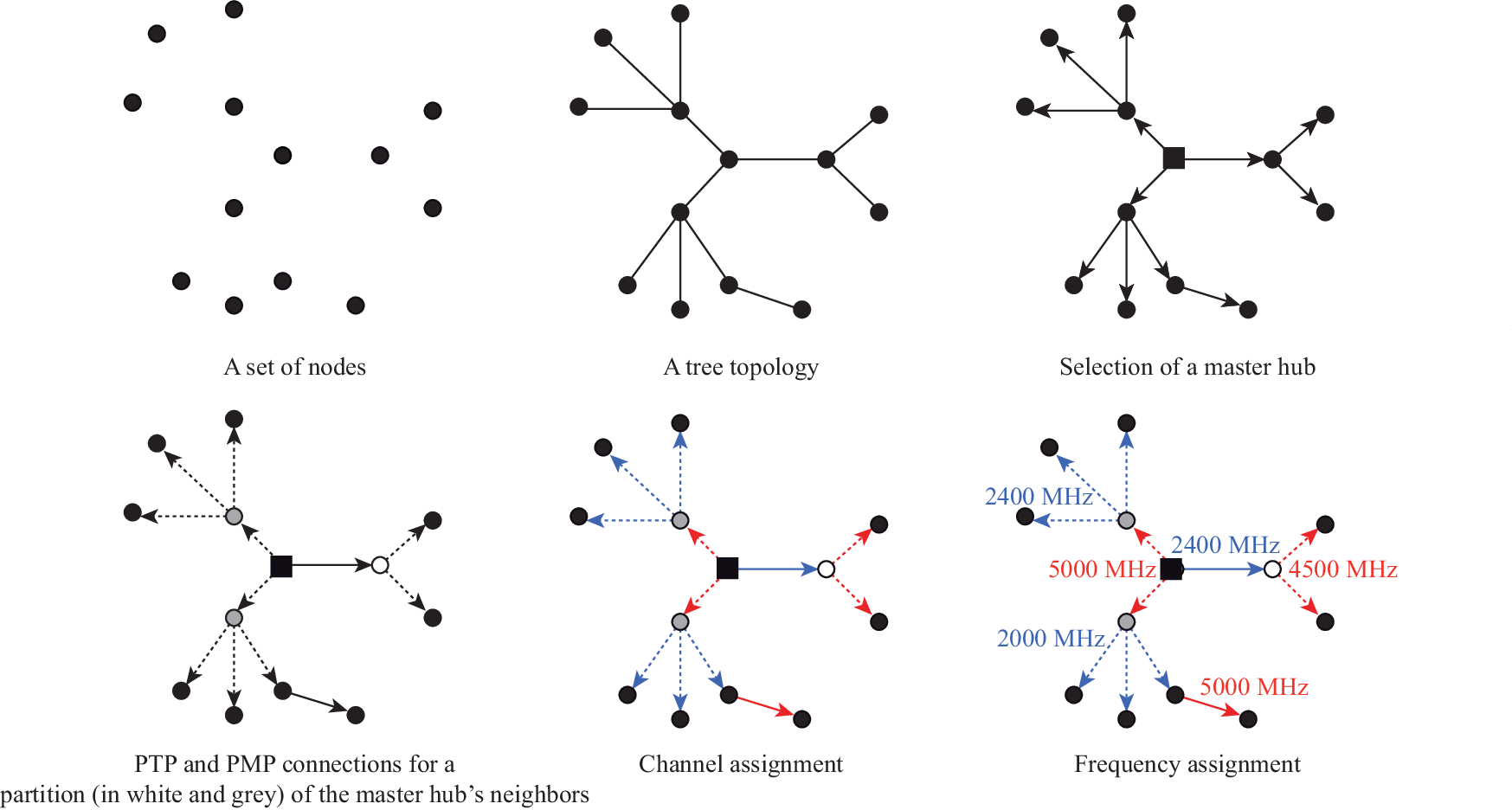}
\caption{Illustration of the network design process (taken from \cite{WH1})}
	\label{fig:process}
\end{figure}

Figure \ref{fig:process}, reproduced from \cite{WH1}, provides an overview of the network design process. Starting from an initial set of nodes, a tree topology $T$ is first constructed to interconnect them. Next, one node is designated as the master hub $r$, depicted as a black square. The three nodes directly connected to this hub are then partitioned into two groups: one group consists of the two grey nodes, while the other contains the white node.
The figure also distinguishes between types of connections: solid lines represent point-to-point (PTP) links, whereas dashed lines correspond to point-to-multipoint (PMP) links. Following this, communication channels are assigned, with one channel illustrated in red and the other in blue. Frequencies are then allocated to each channel; for instance, the red channel is assumed to operate at 4500 MHz and 5000 MHz, while the blue channel operates at 2000 MHz and 2400 MHz.
For the sake of clarity, the figure does not display the activated antenna beams or their precise orientations. These are determined separately through a straightforward geometry-based procedure \cite{2}, which configures each antenna to establish connections between a node and its direct successors.

For every edge $uv$ of the tree topology, the direct throughput $TP_{uv}$ is evaluated by explicitly modeling the physical signal, while also accounting for interference effects between pairs of edges operating on the same frequency. Further details on this computation can be found in \cite{WH1}.

Three different traffic scenarios are considered and defined as follows. Let $n^X_{uv}$ denote the number of data flows associated with scenario $X$ on edge $uv$, and let $d_v$ represent the number of nodes $w$ such that a path exists from $v$ to $w$ in the rooted directed tree.

\begin{itemize}\itemsep=-1pt
\item Scenario $A$: A single data stream exists between any pair of nodes, with only one communication active at a time. In this case, every edge carries exactly one stream, i.e., $n^A_{uv}=1$ for all edges $uv$.

\item Scenario $B$: The master hub communicates simultaneously with all other nodes, resulting in $|V|-1$ concurrent streams. For an edge $uv$, the number of streams depends on its orientation: $n^B_{uv}=d_v$ if the edge is directed from $u$ to $v$, and $n^B_{uv}=d_u$ otherwise.

\item Scenario $C$: Every node exchanges data with every other node, leading to $|V|(|V|-1)$ simultaneous streams. Each edge therefore aggregates both downstream and upstream traffic. The resulting load is given by $n^C_{uv}=2d_v(|V|-d_v)$ if the edge is directed from $u$ to $v$, and $n^C_{uv}=2d_u(|V|-d_u)$ otherwise.
\end{itemize}

The goal is to maximize both the minimum effective throughput, in order to avoid bottlenecks, and the average throughput, so as to ensure a balanced global performance.
Introducing a parameter $p$ to control the trade-off between average and minimum throughput, and a weight $\omega_X$ for each scenario $X \in \{A,B,C\}$, the  function to be maximized is defined as:
$$\sum_{X\in\{A,B,C\} } \omega_X \left(
\min_{[u,v] \in E} \frac{TP_{uv}}{n^X_{uv}}
+ 
p\;\operatorname*{mean}_{[u,v] \in E} \frac{TP_{uv}}{n^X_{uv}}
\right).
$$

The value of this objective function depends on the tree topology, the selected master hub, the partition of the master hub’s successors, the channel assignment, the frequency assignment, the antenna configuration, parameters $p$ and the weigths $\omega_X$ of each scenario $X \in \{A,B,C\}$. The influence of each of these ingredients on the value, quality, and overall performance of the constructed network is precisely the central objective of this paper. 

\section{Litterature review}\label{sec:litterature}

Several studies have already highlighted and clearly demonstrated the important role that certain nodes can play within a network, showing that their position and number of neighbors can have a significant impact on overall network performance, including aspects such as efficiency, stability, and robustness. For example, \citet{wang2016} show that node degree and traffic distribution in wireless sensor networks can significantly influence communication reliability. \citet{wzorek2021router} investigate the structural role of key nodes in emergency and disaster-response communication systems, and demonstrate that the positioning of central nodes strongly affects network reachability, load distribution, and robustness under partial node failures. \citet{shukla2023angle} introduces an angle-based metric for identifying critical nodes in wireless sensor networks. They examine the structural importance of a node through the geometric distribution of its neighboring nodes, and show that some nodes have a big influence on preserving network connectivity and ensuring communication reliability. \citet{khuller1994low} demonstrate how bounding node degree influences network connectivity and resilience in spanning tree design, However, these structural analyses generally neglect important operational aspects of the network, in particular the role of beam control and channel assignment. As a result, they do not capture how communication resources are actually allocated, nor how these mechanisms can influence the effective connectivity and performance of the network.

Studies that incorporate directional antennas typically rely on single-beam devices with discretized orientation. In \cite{hamami}, for instance, each node is equipped with a directional antenna whose beam can be oriented in one of eight possible directions (at $45^\circ$ increments). A communication link is established only when two nodes mutually align their beams in compatible directions. In addition, transmission is carried out over one of two orthogonal channels, assumed to be free of interference. As a result, network connectivity is primarily governed by geometric beam alignment and channel compatibility, without any notion of hierarchical structure induced by a designated master hub, which enforces degree constraints. A related modeling approach is presented in \cite{zhou}, where antennas can be oriented along 6 to 12 discrete directions, and connectivity is defined as a binary function of directional alignment. In this framework, however, no explicit channel or frequency assignment is considered. A similar perspective is adopted in \cite{kumar2006topology}, where each node is equipped with 3 or 4 fixed directional antennas with moderate beamwidths, designed to limit interference while preserving coverage. The resulting topology is approximated using a minimum-degree spanning tree. Although this work moves closer to realistic deployment conditions by combining directional coverage with structural constraints, antennas are still modeled as single-beam interfaces, and neither channel coordination nor hierarchical network organization is considered.

Multi-beam antennas are, for instance, considered in \cite{mumey}, where each antenna is modeled as being composed of $M$ conical sectors, each with an angular width of $360^\circ/M$. These sectors can be activated independently, enabling a single node to support multiple simultaneous directed links. Although this increases connectivity and overall network throughput, it also splits the available transmission power across the active beams, which in turn reduces the gain associated with each individual link. A similar perspective is adopted in \cite{promponas2023optimizing} by considering multi-sector antennas and demonstrating that increasing the number of independently controllable sectors leads to higher network capacity, although the gains tend to diminish as the number of sectors grows. It is also worth mentioning the work of \cite{shi2018modeling} who compare point-to-point (PTP) and point-to-multipoint (PMP) architectures, showing that PMP configurations offer improved spectral efficiency, emphasizing the importance of connection organization rather than antenna characteristics alone.
In these works, multi-beam capability is regarded as a central design parameter. However, the models rely on the assumption of a single communication channel, thereby overlooking issues related to resource coordination and potential interference across multiple channels.

Several studies concentrate on models that jointly consider channel coordination and directional communication.. For example, in \cite{marina} and \cite{zhou2008}, nodes are equipped with single-beam directional antennas characterized by cone-shaped coverage regions, and multiple frequency channels are used to distribute traffic and mitigate local interference. In these approaches, the feasibility of a link depends jointly on beam alignment and channel selection, and allocating communications across different channels contributes to improving overall throughput. Despite relying on similar antenna assumptions, the two studies differ in how they model interactions: 
A weighted conflict metric is used to characterize link interference in \cite{marina}, whereas congestion on heavily loaded links is addressed in \cite{zhou2008}.
 In both cases, however, the network is modeled without the hierarchical structure induced by the designated master hub, which enforces degree constraints.

In summary, existing studies on wireless network modeling adopt a variety of simplifying assumptions, largely driven by the specific application context being addressed. Some approaches emphasize geometric constraints associated with antenna directionality, while others introduce structural restrictions to facilitate connectivity, and many deliberately neglect certain operational parameters in order to preserve computational tractability. As a result, the configuration parameters explored in the literature are diverse and are often examined in isolation rather than in combination. Studies centered on directional antennas often overlook channel assignment issues, research on multi-beam systems typically does not incorporate the hierarchical organization induced by the master hub, and structural analyses of network reliability tend to ignore beam control mechanisms.

Taken together, these works highlight that factors such as the antenna model, the number of active beams, channel assignment, and node degree limitations each have a significant impact on network performance. However, they are rarely integrated within a unified framework. The approach developed in this work builds on these observations by jointly considering all these aspects within a single model, specifically tailored to tree-based tactical wireless networks that include degree-constrained hubs, channel assignment and multi-beam directional communication capabilities.

 \section{Statistical Analysis Approach}
\label{sec:statistical_analysis}

Two algorithms are available to address the problem described in Section \ref{sec:problem}. The first, proposed in \cite{2}, is a parallel Tabu Search variant known as Tabu Beam Search. The second, presented in \cite{WH1}, is a Tabu Search method that integrates several heuristic subroutines. The latter has demonstrated superior performance and will therefore be used to carry out the sensitivity analysis.

The sensitivity analysis is carried out on a set of synthetic benchmark instances specifically generated for this study. Each instance models a tactical wireless network as a connected graph, where nodes represent communication units and edges correspond to potential communication links. The \emph{network size} is defined as the total number of nodes in the graph.

In order to ensure consistency with the baseline configuration and to enable meaningful comparisons, all instances are generated using a uniform procedure with identical parameter settings. Four network sizes are considered, namely 10, 15, 20, and 30 nodes. For each size, ten independent instances are created, leading to a total of forty test instances.

This set of instances is designed to provide a representative range of problem complexities, allowing us to evaluate how the different strategies behave as the network scale increases, and whether their relative performance remains consistent across varying sizes.
Since all strategies are evaluated on exactly the same set of instances, this ensures a fully paired experimental design, which eliminates any bias due to instance-specific variability.

To determine whether the observed variations between different parameter configurations are statistically significant, we adopt a non-parametric statistical testing framework. This choice is motivated by the characteristics of the collected performance data, which do not satisfy the assumption of normality and may also include outliers that could bias classical parametric tests. Furthermore, since all configurations are evaluated on exactly the same set of problem instances, the resulting samples are inherently paired and thus require methods that account for this dependency structure.

The selected statistical tests are based on rank transformations rather than raw performance values, which provides resilience to deviations from distributional assumptions and extreme observations. For these reasons, such approaches are widely used and particularly well suited for comparing the performance of optimization algorithms in experimental studies \cite{demvsar2006}.

\begin{itemize}
  \item \textbf{Global comparison (Friedman test)}: 
   The Friedman test \cite{Friedman1937} is used in order to evaluate whether statistically significant differences exist among the different configurations for a given performance metric. This test is a non-parametric procedure specifically designed for repeated-measures settings, where the same set of problem instances is used to compare multiple configurations.

The methodology consists in ranking the performance of each configuration separately within each instance, assigning rank 1 to the best-performing configuration, rank 2 to the next best, and so on. These instance-wise ranks are then aggregated by computing the average rank of each configuration over all instances. The central question is whether the observed differences in these mean ranks are large enough to be considered statistically significant rather than resulting from random variation.

Formally, the null hypothesis ($H_0$) states that all configurations are equivalent in terms of performance, meaning that any observed differences are due to chance. In contrast, the alternative hypothesis ($H_1$) asserts that at least one configuration performs differently from the others.

To decide between these hypotheses, the Friedman statistic is computed and converted into a $p$-value. The null hypothesis is rejected when this $p$-value is smaller than the chosen significance level, set to $\alpha = 0.05$, indicating that at least one configuration exhibits a statistically significant difference in performance..

  \item \textbf{Post–hoc pairwise comparisons (Wilcoxon signed–rank test)}: 
When the Friedman test reveals the presence of statistically significant differences among the configurations, we carry out a more detailed post-hoc analysis based on pairwise comparisons using the Wilcoxon signed-rank test \cite{wilcoxon1992}. This test is particularly well suited to our setting, as it is a non-parametric procedure designed for paired data, where each pair of configurations is evaluated on exactly the same set of problem instances. Its objective is to determine whether the median of the distribution of performance differences between two configurations can be considered equal to zero, or whether a systematic advantage exists for one configuration over the other.

Because this procedure involves performing multiple pairwise comparisons across all configurations, it is necessary to account for the increased risk of Type I error inflation. To this end, we control the family-wise error rate using a Bonferroni correction. The significance level is therefore adjusted according to the number of pairwise comparisons, yielding:
 \[
    \alpha' = \frac{\alpha}{k(k-1)/2},
  \]
 where $k$ denotes the total number of configurations being compared. Under this corrected threshold, any pair of configurations for which the resulting $p$-value is strictly smaller than $\alpha'$ is regarded as exhibiting a statistically significant difference in performance.
  \end{itemize}

Beyond statistical significance, we also quantify performance differences in relative terms by computing the percentage gain $\Delta\%$ with respect to a baseline configuration, defined as:
\[
\Delta\% = \frac{f(x) - f(x^{\text{base}})}{f(x^{\text{base}})} \times 100,
\]
where $f(x)$ represents the performance achieved under a given configuration, and $f(x^{\text{base}})$ corresponds to the baseline performance.

In addition, median values are reported across instances in order to capture the central tendency of the results and to provide a robust and easily interpretable summary of performance.

Overall, this combined methodology ensures that comparisons between configurations are supported both by rigorous statistical testing and by meaningful performance indicators, allowing for a comprehensive and reliable evaluation of the results.

\section{Sensitivity analysis of master hub selection}\label{sA:masterhub}

In the considered tactical network model, the master hub serves as the central control node from which all communications are coordinated. In principle, any node in the network may be selected to assume this role. Once designated, the master hub becomes the root of the topology, inducing a directed tree structure (arborescence) in which all edges are oriented outward from this central node. This orientation defines the hierarchical organization of the network and determines the direction in which information flows between nodes.

The selection of the master hub is therefore a fundamental design decision, as it has a direct impact on routing paths, interference interactions, and, more generally, on the global performance of the network. In the baseline setting adopted in \cite{WH1}, the master hub is chosen using a predefined selection rule that combines node degree and eccentricity. This rule aims to identify nodes that are both well connected (high degree) and topologically central (low eccentricity), thereby achieving a balance between local connectivity and global positioning within the network.

To better understand the role of this design choice, we conduct a sensitivity analysis on the decision variable associated with master hub selection. The goal is to evaluate how alternative hub-selection strategies influence both the resulting network performance and the behavior of the Tabu Search optimization process.

\subsection{Master Hub Selection Strategies}\label{sec:hubstrategies}

To investigate the impact of the master hub selection on network behavior, several selection strategies have been designed and compared. Each of them reflects a different structural assumption regarding which node characteristics are most conducive to efficient network coordination. The underlying objective is to determine whether performance is primarily driven by local connectivity, global centrality, or a trade-off between the two.

The baseline strategy follows the master hub selection procedure adapted from \cite{WH1} and is used as the reference processing mode. In a first step, all leaf nodes of the topology are removed from the candidate list, since peripheral nodes are generally less suitable candidates for acting as the central coordination point of the network.
The remaining nodes are then evaluated according to two structural indicators: the node degree, which captures local connectivity, and the eccentricity, which reflects the node’s position within the global topology. Each candidate is ranked with respect to these two measures, and the resulting rankings are aggregated into a single composite score.
Nodes whose combined score exceeds the median (50th percentile) are retained, leading to a reduced candidate set that filters out both highly peripheral nodes and nodes that are excessively central, which may not efficiently support traffic distribution in a balanced manner. 

This baseline procedure denoted \texttt{baseline} serves as the reference configuration against which all alternative selection strategies are evaluated. The alternative strategies are defined as follows:

\begin{itemize}

\item Leaf-only (\texttt{LeafOnly}): this strategy restricts the candidate set exclusively to leaf nodes, thereby exploring the extreme case in which the master hub is located at the periphery of the network topology.

\item Leaf-and-non-leaf (\texttt{LeafNonLeaf}): this configuration combines the preprocessing-based selection (derived from degree and eccentricity ranking) with an explicit inclusion of all leaf nodes. More precisely, it merges the set of nodes obtained from the baseline preprocessing stage with the set of leaf nodes in the topology. This hybrid design allows the exploration of both central and peripheral candidates while excluding intermediate nodes, thereby testing whether extreme structural positions are more favorable for the role of master hub.

\item All-nodes (\texttt{AllNodes}): in this strategy, every node in the network is considered as a valid candidate for the master hub role. This represents a neutral configuration with no structural filtering or bias, allowing the optimization process to freely explore all possible hub locations.

\item Max-degree (\texttt{MaxD}): this strategy selects nodes with the highest degree, corresponding to a purely local criterion. It is based on the assumption that nodes with the largest number of direct connections are better suited to act as coordination points.

\item Min-eccentricity (\texttt{MinE}): this approach selects nodes with the smallest eccentricity, thus favoring a global centrality criterion. It prioritizes nodes that minimize the maximum shortest-path distance to all other nodes in the network.

\item Weighted Centrality–Connectivity (\texttt{70E30D} and (\texttt{30E70D}): in addition to the previous rules, we consider a weighted scoring strategy that combines degree and eccentricity in order to evaluate the trade-off between local connectivity and global centrality. Two parameterizations are tested: one emphasizing eccentricity (70\% eccentricity / 30\% degree), and another emphasizing connectivity (30\% eccentricity / 70\% degree). This allows us to investigate whether the master hub is more effective when it is structurally central or when it is highly connected.

\end{itemize}

Each of these strategies was integrated within the Tabu Search optimization framework of \cite{WH1} under identical experimental conditions, including runtime limits, instance sizes, and stopping criteria.
This systematic comparison provides a structured basis for evaluating how topological properties influence the search dynamics and overall performance of the tactical network design process.

\subsection{Evaluation Metrics and Statistical Analysis}

\subsubsection{Evaluation Metrics}

To quantitatively evaluate the impact of each master hub selection strategy on network performance, we compute a set of five metrics for all instances and configurations. These indicators are designed to capture both the quality of the solutions produced and the behavior of the Tabu Search optimization process.

The following performance measures are considered:

\begin{itemize}

\item Best Objective Value: this corresponds to the best value of the objective function obtained during a given Tabu Search run. It serves as the primary indicator of solution quality and overall network performance.

\item Iteration Found: denotes the iteration index at which the best objective value is first reached. This metric provides information about the convergence speed of the algorithm.

\item Time Found: represents the elapsed computational time required to reach the best solution. It reflects the temporal efficiency of the search process.

\item Total Iterations: indicates the total number of iterations executed before termination of the algorithm. This measure captures the overall exploration effort and provides insight into the stability and duration of the search.

\item Time per Iteration: measures the average computational time (in milliseconds) required for a single iteration, thereby quantifying the computational cost associated with each configuration.

\end{itemize}

\subsection {Experimental Results}

Table~\ref{tab:stat_tests_full} summarizes the results of the experiment.
For every metric, we first report the outcome of the Friedman test (column Ftest), where Y indicates a statistically significant difference between the strategies and N indicates a non-significant result. We also report the median performance value (column MedianB)  achieved by the \texttt{baseline} strategy. Since the Friedman test reveals a significant effect for all metrics, we proceed with pairwise comparisons between each strategy $S$ and \texttt{baseline} using the Wilcoxon test (column Wtest), with Y and N again denoting statistically significant and non-significant differences, respectively. We also report the median performance value obtained by $S$ (column MedianS). In cases where the Wilcoxon test indicates a significant difference between $S$ and \texttt{baseline}, we further identify the method achieving the best performance (column Best).


\begin{table}[H]
\centering
\scriptsize
\begin{tabular}{lcccrrc}

Metric & Ftest & WTest&  Strategy & MedianB & MedianS & Best  \\
\toprule
\midrule
Best Objective Value & Y &   &  &  33.2&  &   \\
    &       & Y     &\texttt{LeafOnly} && 27.7 &  \texttt{baseline}  \\
 	&		&	N	&	\texttt{AllNodes	}&&	 33.2 	&	 		\\
 	&		&	N	&	\texttt{LeafNonLeaf	}&&	 33.2 	&	 	\\
 	&		&	N	&	\texttt{MaxD	}&&	 33.9 	&	 	\\
 	&		&	N	&	\texttt{30E70D}	&&	 33.2 	&	 	\\
 	&		&	N	&	\texttt{70E30D}&&	 33.4 	&	 		\\
 	&		&	N	&	\texttt{minE}	&&	33.0	&	 		\\
\midrule
Iteration Found 	& Y &   &  &  64&  &		\\
 	&		&	Y	&	\texttt{LeafOnly}	&&	136	&		\texttt{baseline}	\\
 	&		&	N	&	\texttt{AllNodes	}&&	50	&			\\
 	&		&	N	&	\texttt{LeafNonLeaf}	&&	38	&			\\
 	&		&	Y	&	\texttt{MaxD	}&&	130	&		\texttt{baseline}	\\
 	&		&	N	&	\texttt{30E70D	}&&	64	&			\\
 	&		&	N	&	\texttt{70E30D}	&&	68	&			\\
 	&		&	N	&	\texttt{minE	}&&	81	&			\\
\midrule
Time Found 	& Y &   &  &  1195&  &		\\
 	&		&	N	&	\texttt{LeafOnly	}&&	684	&	 		\\
 	&		&	Y	&	\texttt{AllNodes}	&&	 1523 	&	 	\texttt{baseline}	\\
 	&		&	N	&	\texttt{LeafNonLeaf	}&&	 1448 	&	 		\\
 	&		&	N	&	\texttt{MaxD	}&&	 1182 	&	 	\\
 	&		&	N	&	\texttt{30E70D}	&&	 1198 	&	 		\\
 	&		&	Y	&	\texttt{70E30D	}&&	 1079 	&	 	\texttt{70E30D	}\\
 	&		&	N	&	\texttt{minE	}&&	759	&	 		\\
\midrule
Total Iterations 	& Y &   &  &  1110&  &	\\
 	&		&	Y	&	\texttt{LeafOnly	}&	&  4713		  	&	\texttt{LeafOnly}	\\
 	&		&	Y	&	\texttt{AllNodes}	&	& 822		  	&	\texttt{baseline}	\\
 	&		&	Y	&	\texttt{LeafNonLeaf}	&	& 716		  	&	\texttt{baseline}	\\
 	&		&	Y	&	\texttt{MaxD}	&	 & 1918 		  	&	\texttt{MaxD	}\\
 	&		&	Y	&	\texttt{30E70D	}&	 &  1149 		  	&	\texttt{30E70D	}\\
 	&		&	Y	&	\texttt{70E30D	}&	 & 1182		  	&	\texttt{70E30D	}\\
 	&		&	Y	&	\texttt{minE	}&	 & 2014 		  	&	\texttt{minE	}\\
\midrule
Time per Iteration  	& Y &   &  & 5179 &  &		\\
 	&		&	Y	&	\texttt{LeafOnly}	&&	 1.281 	&	 	\texttt{LeafOnly	}\\
 	&		&	Y	&	\texttt{AllNodes}	&&	 6.876 	&	 	\texttt{baseline}	\\
 	&		&	Y	&	\texttt{LeafNonLeaf}	&&	 6.981 	&	 	\texttt{baseline}	\\
 	&		&	Y	&	\texttt{MaxD	}&&	 2.779 	&	 	\texttt{MaxD	}\\
 	&		&	Y	&	\texttt{30E70D}&&	 4.915 	&	 	\texttt{30E70D}	\\
 	&		&	Y	&	\texttt{70E30D	}&&	 4.783 	&	 	\texttt{70E30D}\\
 	&		&	Y	&	\texttt{minE	}&&	 2.561 	&	 	\texttt{minE	}\\
\bottomrule
\end{tabular}

\caption{Statistical comparison of master hub strategies.
\label{tab:stat_tests_full}
}
\end{table}

\subsection{Visualization and interpretation}
To complement the statistical analysis, we present boxplots illustrating the distribution of each performance metric across all master hub strategies. These visualizations emphasize the median values, the dispersion of results, and any potential outliers, offering an intuitive way to compare the variability and consistency of each strategy.

\begin{figure}[H]
\centering

\begin{subfigure}{0.48\textwidth}
    \centering
    \includegraphics[width=\linewidth]{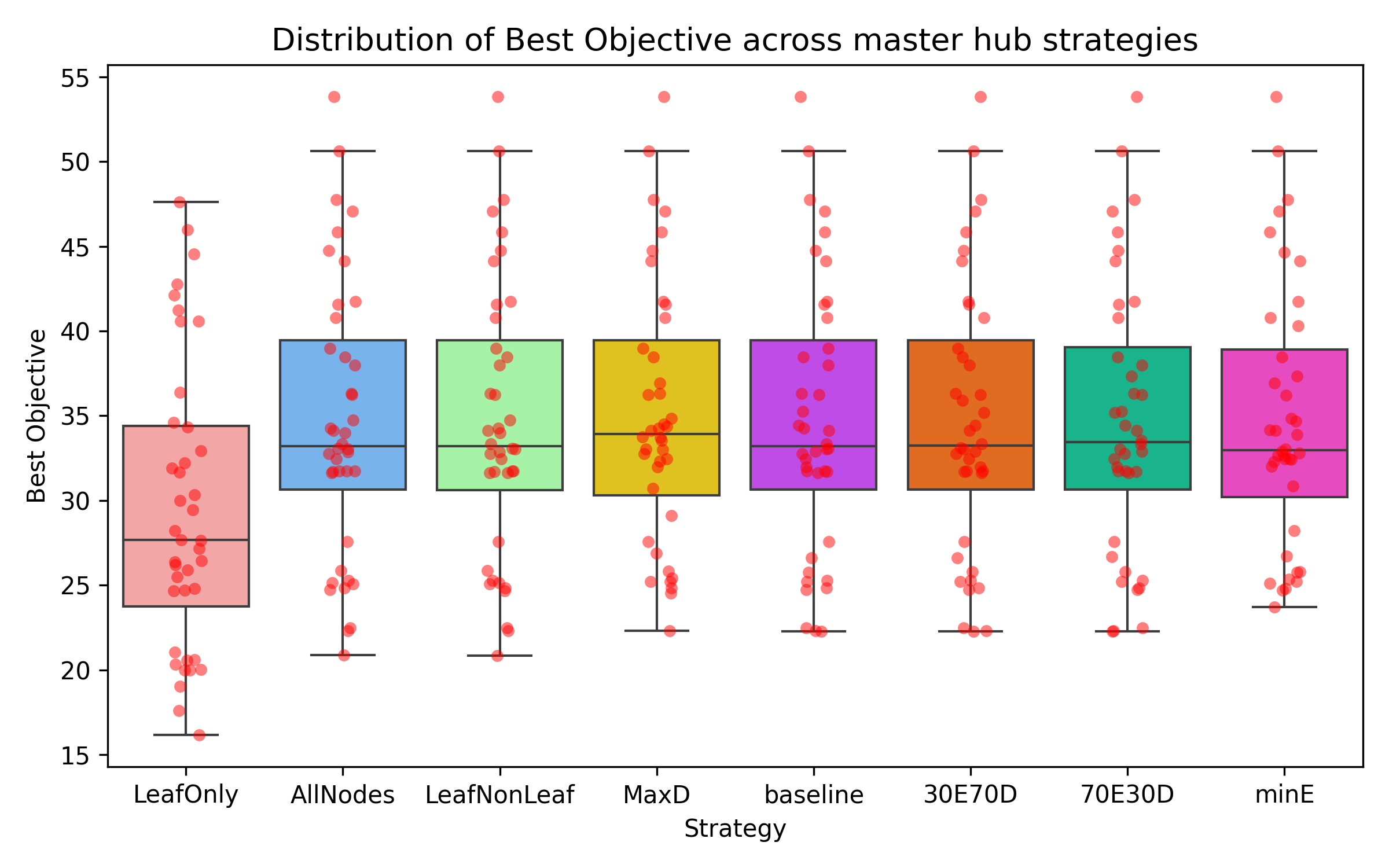}
    \caption{Best Objective Value}
    \label{fig:box_obj}
\end{subfigure}
\hfill
\begin{subfigure}{0.48\textwidth}
    \centering
    \includegraphics[width=\linewidth]{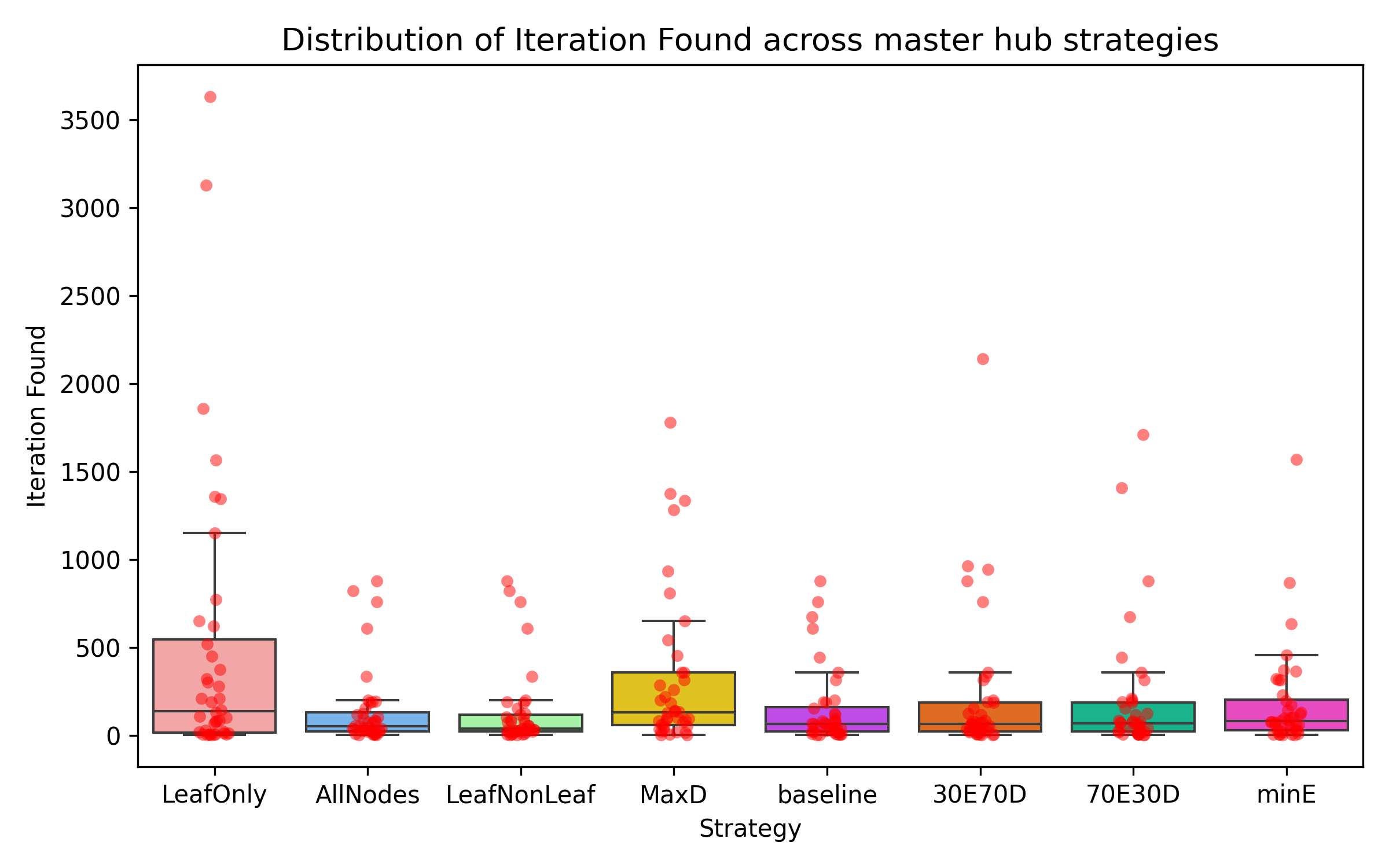}
    \caption{Iteration Found}
    \label{fig:box_iterfound}
\end{subfigure}

\vspace{0.5cm}

\begin{subfigure}{0.48\textwidth}
    \centering
    \includegraphics[width=\linewidth]{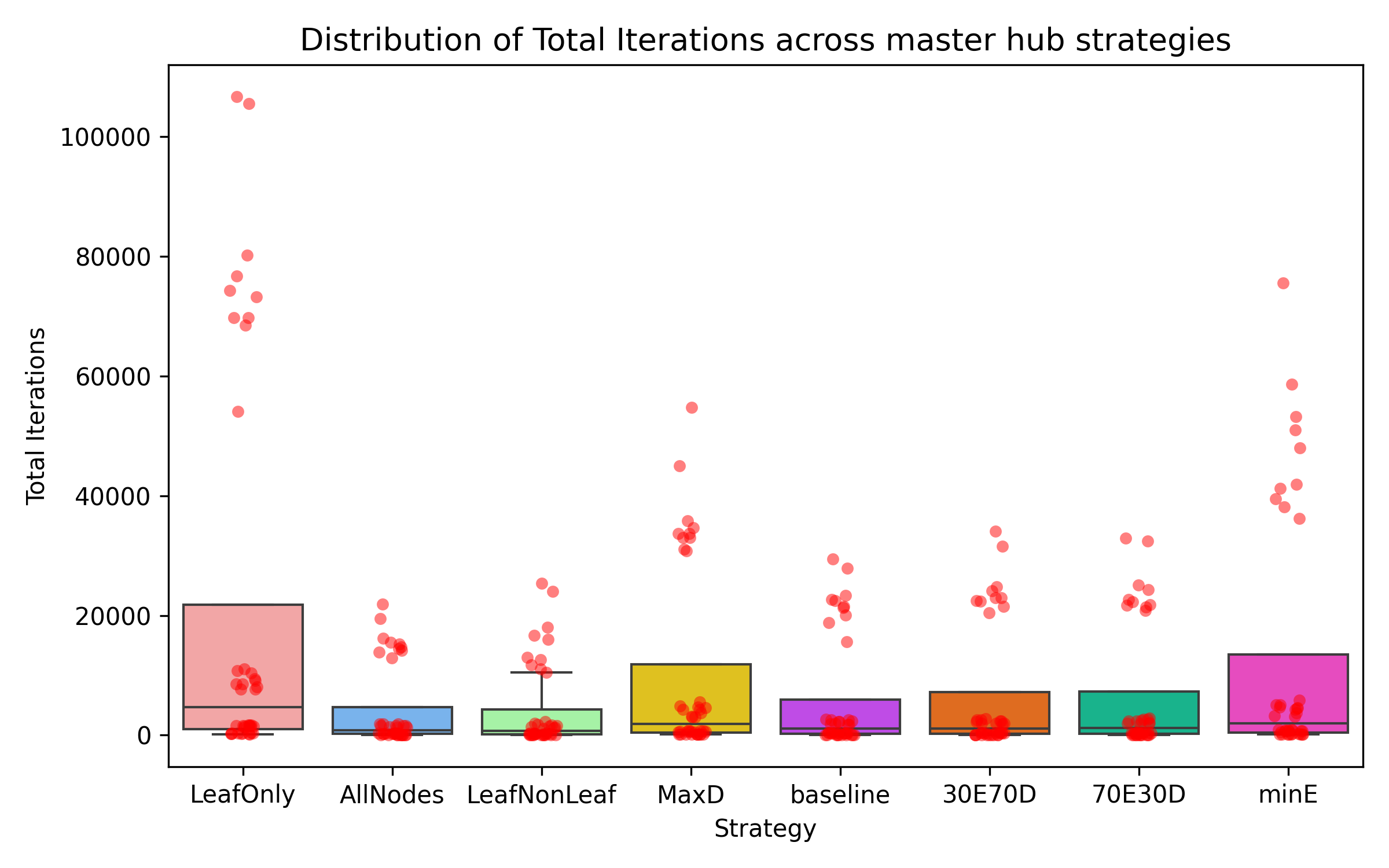}
    \caption{Total Iterations}
    \label{fig:box_totalit}
\end{subfigure}
\hfill
\begin{subfigure}{0.48\textwidth}
    \centering
    \includegraphics[width=\linewidth]{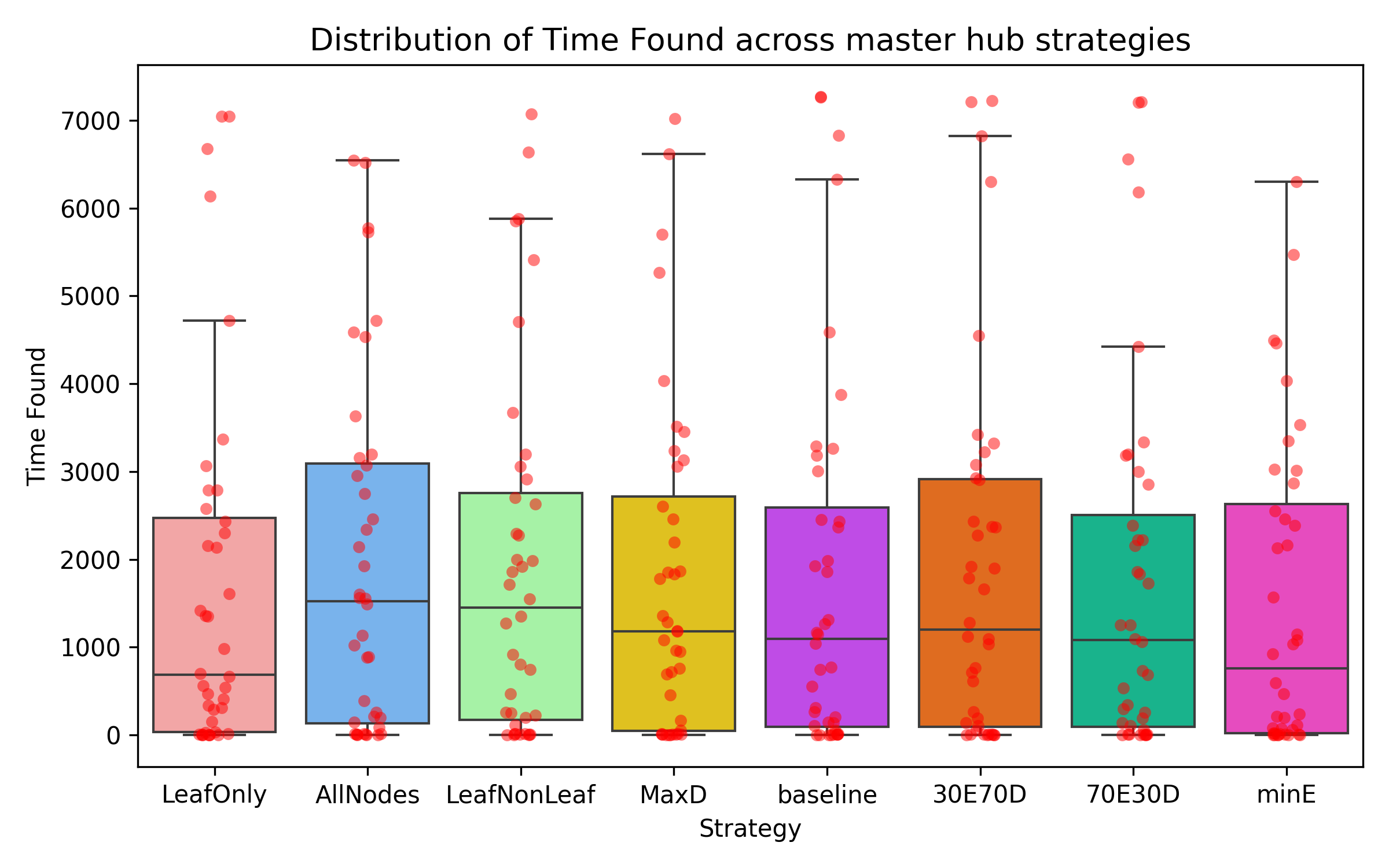}
    \caption{Time Found (s)}
    \label{fig:box_timefound}
\end{subfigure}

\caption{Comparison of network performance metrics across master hub selection strategies.}
\label{fig:all_boxplots}
\end{figure}

The boxplots in Figure~\ref{fig:all_boxplots} offer a complementary perspective on performance variability across master hub strategies. For the Best Objective Value (Figure~\ref{fig:all_boxplots}a), all strategies, except \texttt{LeafOnly}, exhibit similar median performance. The markedly lower median and greater dispersion observed for \texttt{LeafOnly} support the hypothesis that placing the master hub at the network periphery is structurally disadvantageous.

By contrast, strategies that combine degree and eccentricity, including the \texttt{baseline} strategy and its two weighted variants, consistently yield stable objective values, suggesting that an effective master hub should be both highly connected and centrally located. Notably, the \texttt{MinE} strategy produces a slightly lower objective value than \texttt{baseline}, indicating that eccentricity alone, while useful for identifying central nodes, does not fully capture the connectivity required to maximize throughput.

For Iteration Found (Figure~\ref{fig:all_boxplots}b), the \texttt{baseline} strategy reaches the best solution in fewer iterations than the weighted variants \texttt{30E70D} and \texttt{70E30D}, as well as \texttt{MaxD} and \texttt{MinE}. This indicates that relying solely on degree or applying extreme weightings can slow the search by directing it toward less relevant candidate hubs.
Although \texttt{AllNodes} and \texttt{LeafNonLeaf} occasionally converge more quickly, this is largely due to the evaluation of a broader set of potential hubs, increasing search coverage rather than true efficiency. As expected, \texttt{LeafOnly} also requires more iterations, since it confines the search to structurally weak candidates.

A similar pattern is evident for Total Iterations (Figure~\ref{fig:all_boxplots}c) and Time Found (Figure~\ref{fig:all_boxplots}d). Strategies that evaluate all nodes, such as \texttt{AllNodes}, incur higher computational costs because they explore a larger set of hub configurations. In contrast, \texttt{LeafOnly} appears faster, but this apparent efficiency is misleading, as it converges to low-quality solutions.

Overall, the \texttt{baseline} strategy achieves a particularly effective balance between exploration and exploitation. On the one hand, it restricts the search space by focusing on structurally relevant candidates, thereby avoiding unnecessary evaluations. On the other hand, it preserves sufficient diversity among the considered solutions, which helps prevent premature convergence to local optima. As a result of this balance, the approach typically requires fewer iterations, reduces overall computation time, and maintains consistently stable and reliable solution quality across all tested instances.

\section{Sensitivity analysis of PMP successor limits}

In the baseline configuration, the maximum number of PMP successors per relay node is set to ten. This constraint also indirectly limits the maximum degree of the master hub, as it can serve up to two PMP groups. Such a modeling choice plays a key role in shaping how the tree structure develops and how communication resources such as beams, channels, and frequencies are allocated throughout the network. Consequently, the PMP parameter has a direct impact on the depth of the topology, the connectivity of the master hub, interference patterns, and, ultimately, the overall network throughput.

To assess the influence of this assumption, we perform a sensitivity analysis by varying the maximum number of PMP successors. More precisely, we examine six alternative configurations in which each relay node can support up to 5, 8, 12, 15, 20, or 25 outgoing PMP connections, while keeping all other system parameters unchanged (including the master hub selection strategy, number of channels and frequencies, antenna characteristics, and waveform structure).

For each feasible configuration, the same optimization procedure is applied, and the resulting solutions are compared using the objective function as the main performance indicator.


To ensure a fair and consistent comparison across different PMP successor limits, the set of test instances is carefully selected to guarantee feasibility under each constraint.
For instance, when the maximum number of PMP successors is set to 5, instances with 10 nodes or more can still produce feasible topologies, allowing the full set of graphs of size 10 and above to be considered. In contrast, when this limit is increased to 15, only instances with at least 15 nodes can effectively exploit the relaxed constraint; as a result, the evaluation is restricted to graphs of size 15 and larger. More generally, for an instance of size $n$, only PMP successor limits that do not exceed $n$ are taken into account. Consequently, each experiment is carried out on the subset of instances for which the corresponding structural constraint is both admissible and meaningful.

Following the evaluation protocol described in Section~\ref{sec:statistical_analysis}, performance differences are analyzed using non-parametric statistical tests. For each network size, a global comparison is first performed, followed by pairwise comparisons against a fixed baseline configuration (PMP = 10).


\begin{table}[H]
\centering
\scriptsize
\caption{Sensitivity analysis of PMP successor limits.}
\label{tab:pmp_sensitivity_by_size}
\small
\setlength{\tabcolsep}{4pt}
\begin{tabular}{ccccrrrc}
\toprule
Network size	&	Ftest	&	Wtest	&	PMP	&MedianB	&	MedianS	&	 $\Delta\%$ 	&	Best	 \\
\midrule													
\addlinespace[2pt]												
 	\textbf{10}&	N	&		&		&		&&		&		 \\
\addlinespace[4pt]												
\midrule														
\addlinespace[2pt]												
 	\textbf{15}&	N	&		&		&		&&		&		 \\
\addlinespace[4pt]												
\midrule													
\addlinespace[2pt]	
&	Y	&		&		&	32.22&	&	&		 \\
 	&		&	Y	&	5&  	&	 29.84 	&	 -6.08 	&	baseline	 \\
 	\textbf{20}&		&	N	&8&	  	&	 32.22 	&	 0.00 	&		 \\
 	&		&	N	&	12&  	&	 32.22 	&	 0.00 	&		 \\
 	&		&	N	&	 15& 	&	 32.22 	&	 0.00 	&		 \\
 	&		&	N	&	 20& 	&	 32.22 	&	 0.00 	&		 \\
\addlinespace[4pt]												
\midrule													
\addlinespace[2pt]												
 	&	Y	&		&		&24.98 &		&		&		 \\
 	&		&	Y	&	5& 	&	 21.86 	&	 -11.87 	&	baseline	 \\
	&		&	N	&	 8& 	&	 24.98 	&	 0.00 	&		 \\
	\textbf{30}&		&	N	&	12&  	&	 24.98 	&	 0.00 	&		 \\
 	&		&	N	&	15&  	&	 24.98 	&	 0.00 	&		 \\
 	&		&	N	&	 20& 	&	 24.98 	&	 0.00 	&		 \\
 	&		&	N	&	 25& 	&	 24.98 	&	 0.00 	&		 \\
													
\addlinespace[4pt]												
\midrule													
\addlinespace[2pt]												
 	&	Y	&		&		&33.22&		&		&		 \\
  	&		&	Y	&	 5& 	&	 32.79 	&	 -2.51 	&	baseline	 \\
 	&		&	N	&	 8& 	&	 33.22 	&	 0.00 	&		 \\
 	\textbf{GLOBAL}&		&	N	&	12&  	&	 33.22 	&	 0.00 	&		 \\
	&		&	N	&	 15& 	&	 33.22 	&	 0.00 	&		 \\
 	&		&	N	&	 20& 	&	 33.22 	&	 0.00 	&		 \\
  	&		&	N	&	 25& 	&	 33.22 	&	 0.00 	&		 \\

\midrule
\end{tabular}
\end{table}

Table~\ref{tab:pmp_sensitivity_by_size} shows that the influence of the PMP parameter is highly dependent on the instance size, with clearly different behaviours observed across problem scales. For instances of size 10 and 15, the Friedman test does not reveal any statistically significant differences between the considered settings. Specifically, for size 10, no significant effect is observed when comparing PMP limits of 5, 8 and 10 successors, while for size 15, the same conclusion holds for PMP limits of 5, 8, 10, 12, and 15 successors.
This suggests that the PMP constraint is not binding in this case.

For instances of size 20 and 30, a clear threshold effect emerges. Very restrictive PMP values, in particular PMP = 5, lead to a noticeable degradation in solution quality, with median relative losses of about $6.08\%$ for size 20 and $11.87\%$ for size 30 compared to the baseline PMP = 10; these differences are statistically significant according to the Wilcoxon tests. Increasing the limit from 5 to 8 successors yields a marked improvement, indicating that a minimal level of branching is necessary to capture effective network structures. However, once PMP reaches 10 or higher, the median relative gain $\Delta\%$ becomes zero and all pairwise comparisons are non-significant, showing that the objective values stabilize and that further increases in the number of successors no longer provide meaningful gains. For size 30, a modest improvement is still observed between PMP = 5 and PMP = 8, but this effect remains limited and vanishes entirely beyond PMP = 10. The global results aggregated over the 40 instances are reported at the end of the table.

Overall, these results confirm that PMP = 10 appears to be an appropriate and well-balanced threshold for the considered settings. At this level, the constraint is sufficiently relaxed to avoid any degradation in performance for medium-sized instances, ensuring that the optimization process can fully exploit the relevant network structures. At the same time, increasing the PMP limit beyond 10 does not lead to any statistically significant improvements, nor does it produce noticeable gains in practice. This suggests that higher values only increase flexibility without translating into better objective outcomes.

\section{Sensitivity Analysis on Antenna Beam Width}

We investigate the sensitivity of the proposed approach with respect to the antenna beam width. This parameter governs the angular spread of each transmitted beam and therefore directly influences the trade-off between coverage and directionality. Narrow beam widths concentrate signal energy in a specific direction, resulting in higher antenna gain and reduced interference, but they limit spatial coverage. In contrast, wider beam widths provide broader angular coverage at the expense of lower gain and potentially increased interference. Since beam width affects signal strength, Signal-to-Interference-plus-Noise Ratio (SINR), and spatial reuse, it is a key factor in the performance of the tactical wireless network.

In our experiments, we consider four discrete antenna configurations corresponding to beam widths of 4, 6, 12, and 24 beams. These values are representative of practical antenna technologies and allow us to explore a broad range of angular resolutions, from highly coarse (4 beams) to highly directional (24 beams). For all configurations, we keep the antenna radiation patterns and propagation model unchanged, modifying only the number of beams.

Furthermore, for the 24-beam configuration, we distinguish between two operating modes. The first, referred to as \emph{standard}, assumes that the antenna always operates using directional beams. The second which is the baseline configuration implemented in \cite{WH1} allows the antenna to switch to an omnidirectional radiation pattern when the number of active beams exceeds 7, in line with the specifications provided by our industrial partner. This mechanism is intended to capture realistic hardware constraints and energy limitations in operational settings.

By comparing these configurations, we aim to quantify the impact of beam width on overall network performance and to assess whether the introduction of an omnidirectional fallback, motivated by industrial requirements, induces any significant performance degradation.


The statistical analysis reported in Table~\ref{tab:antenna_sensitivity_by_size_global} indicates that antenna beam width has no statistically significant impact on network performance for small instances. For Size = 10, the Friedman test reveals no statistically significant differences. This can be attributed to the low density of links and short communication distances in small networks, where interference remains minimal and even wide beams provide sufficient coverage and signal quality.

For Size = 15, the Friedman test becomes significant, suggesting that antenna configuration begins to influence performance at a global level. However, none of the pairwise Wilcoxon comparisons remains significant after Bonferroni correction, and the observed performance differences relative to the baseline remain limited (approximately between $-2\%$ and $-6\%$). This reflects a transitional regime in which the effect of beam width starts to emerge, but is still not strong enough to generate clear and consistent pairwise separations. At this scale, antenna directivity becomes relevant, yet interference is not sufficiently dominant for highly directional configurations to provide a clear advantage.

This behavior changes as the network size increases. For Size = 20, both the Friedman test and all pairwise comparisons are statistically significant, and all lower-directivity configurations (4, 6, and 12 beams) are clearly outperformed by the 24-beam OmniSwitch baseline, with performance losses ranging from approximately $-12\%$ to $-19\%$. This indicates that interference becomes the dominant limiting factor as the network grows. In this regime, narrower beams better concentrate transmitted energy toward intended receivers, increasing antenna gain and substantially reducing inter-link interference, which improves SINR and overall network performance.

For Size = 30, the Friedman test remains significant, confirming the strong overall effect of antenna configuration. The pairwise analysis further shows that the 4-beam and 12-beam configurations are significantly worse than the baseline, with large performance degradations reaching up to $-36\%$. Although the difference with the 6-beam configuration is not statistically significant after correction, the observed gap remains substantial (around $-31\%$), which suggests that the lack of significance is more likely due to limited statistical power than to true equivalence.

From a technological standpoint, these results highlight the trade-off between interference mitigation and connectivity. As network size increases, interference becomes the dominant performance bottleneck. Configurations with insufficient directivity (4, 6, and 12 beams) fail to adequately suppress interference, leading to substantial performance losses. In contrast, highly directional configurations (24 beams) effectively limit interference and increase antenna gain. Although narrower beams impose stricter alignment requirements and may reduce connectivity, the benefits in interference reduction dominate in larger networks, which explains the superior performance of the 24-beam configurations.

\begin{table}[H]
\centering
\scriptsize
\caption{Sensitivity analysis of antenna beam width.}
\label{tab:antenna_sensitivity_by_size_global}
\setlength{\tabcolsep}{4pt}
\begin{tabular}{ccccrrcc}
\toprule
Network size	&	 Ftest 	&	Wtest	&	Number of beams 	&	MedianB	&	MedianS	&	 $\Delta\%$ 	&	Best	  \\
\midrule															
\addlinespace[2pt]															
\textbf{10}	&	N	&	  	&		&	 - 	&	 - 	&	 - 	&		  \\
\addlinespace[4pt]															
\midrule															
\addlinespace[2pt]															
 	&	Y	&		&		&	 35.49 	&		&		&		  \\
 	&		&	N	&	4	&		&	 34.92 	&	 -5.65 	&		  \\
\textbf{15}	&		&	N	&	6	&		&	 35.21 	&	 -2.05 	&		  \\
 	&		&	N	&	12	&		&	 34.96 	&	 -3.41 	&		  \\
	&		&	N	&	24 standard	&		&	 35.49 	&	 0.00 	&		  \\
\addlinespace[4pt]															
\midrule															
\addlinespace[2pt]															
	&	Y	&		&		&	 32.07 	&		&		&		  \\
 	&		&	Y	&	4	&		&	 26.04 	&	 -18.79 	&	baseline	  \\
\textbf{20}	&		&	Y	&	6	&		&	 27.73 	&	 -14.02 	&	baseline	  \\
 	&		&	Y	&	12	&		&	 28.71 	&	 -12.45 	&	baseline	  \\
 	&		&	N	&	24 standard	&		&	 32.07 	&	 0.00 	&		  \\
\addlinespace[4pt]															
\midrule															
\addlinespace[2pt]															
 	&	Y	&		&		&	 24.98 	&		&		&		  \\
 	&		&	Y	&	4	&		&	 15.59 	&	 -36.81 	&	baseline	  \\
\textbf{30}	&		&	N	&	6	&		&	 17.45 	&	 -31.47 	&		  \\
 	&		&	Y	&	12	&		&	 17.26 	&	 -22.30 	&	baseline	  \\
 	&		&	N	&	24 standard	&		&	 25.13 	&	 0.00 	&		  \\
\addlinespace[4pt]															
\midrule															
\addlinespace[2pt]															
  	&	Y	&		&		&	 33.22 	&		&		&		  \\
 	&		&	Y	&	4	&		&	 29.69 	&	 -11.07 	&	baseline	  \\
\textbf{GLOBAL}	&		&	Y	&	6	&		&	 31.09 	&	 -5.40 	&	baseline	  \\
  	&		&	Y	&	12	&		&	 31.48 	&	 -5.41 	&	baseline	  \\
  	&		&	N	&	24 standard	&		&	 33.22 	&	 0.00 	&		  \\
    \bottomrule
\end{tabular}

\end{table}

\section{Sensitivity Analysis of Antenna Technology}

We investigate the impact of antenna technology on network performance by comparing single-beam and multi-beam configurations. A single-beam antenna can transmit in only one direction at a time, which restricts simultaneous communications and introduces additional scheduling constraints. In contrast, multi-beam antennas enable concurrent transmissions toward multiple neighbors, making them particularly well suited for PMP communications.

Beyond the aggregated objective value, we also record the best topology obtained for each individual traffic scenario $X \in {A,B,C}$ (see Section \ref{sec:problem}). This finer-grained analysis allows us to go beyond global performance indicators and to isolate the influence of antenna technology across different traffic patterns and network scales.

The results reported in Table~\ref{tab:single_vs_multi} reveal several clear structural trends. For small and medium-sized instances (10, 15, and 20 nodes), single-beam antennas consistently achieve higher overall objective values in most runs. This effect is particularly pronounced for networks of size 10 and 15, where single-beam configurations dominate both in terms of aggregated objective and across nearly all individual traffic scenarios. These observations suggest that, in smaller networks, interference levels remain sufficiently low that the additional spatial flexibility offered by multi-beam antennas does not translate into performance gains.
For networks of size 20, the gap between the two technologies begins to narrow, although single-beam antennas still outperform multi-beam configurations in most cases. For instance, in Instance~1, the overall objective reaches $45.8989$ with single-beam antennas compared to $33.3545$ for multi-beam, while in Instance~7 the values are $37.428$ versus $31.6522$, respectively. However, this dominance is no longer systematic: in Instance~4, the results are very close ($32.8353$ vs $32.4574$), indicating that multi-beam configurations are becoming increasingly competitive.

A scenario-level breakdown further highlights this transition. In Scenario~$B$ (traffic concentrated around the master hub), for example, Instance~2 shows a clear advantage for multi-beam antennas ($27.7807$ vs $16.8122$), and a similar pattern is observed in Instance~6. This suggests that, as the network grows, the ability to support simultaneous directional transmissions helps alleviate congestion around central nodes. Comparable conclusions can be drawn for Scenario $A$ and $C$.

The transition becomes even more evident for networks of size 30. While single-beam antennas still perform well in some cases (e.g., Instance~5 with $32.0735$ compared to $24.7545$ for multi-beam), several instances exhibit severe performance degradation. In Instance~6, the overall objective drops to $2.3452$ for single-beam versus $25.2939$ for multi-beam, and a similar pattern is observed in Instance~8 ($2.3257$ vs $22.5007$).
These large discrepancies reveal a structural instability of single-beam configurations in larger networks. As interference accumulates and the topology becomes more complex, the inability to support simultaneous directional transmissions leads to significant throughput bottlenecks. Multi-beam antennas, by contrast, maintain much more stable performance levels, typically ranging between $22$ and $27$ across instances.
This behavior suggests the emergence of a transition regime around networks of size 30, beyond which multi-beam technology provides clear structural resilience against interference and connectivity degradation.

From a scenario perspective, Scenario~$A$ (single indirect communication) tends to favor single-beam antennas for small and medium instances, where long PTP chains can effectively optimize throughput. Scenario~$B$ (all nodes communicating with the master hub) increasingly benefits from multi-beam capabilities as network size grows, due to traffic concentration around the hub. Scenario~$C$ (all-to-all communication) exhibits intermediate behavior, with multi-beam advantages becoming more apparent only at larger scales.

Overall, the results show that the benefit of multi-beam antennas is strongly dependent on network size. While single-beam technology is sufficient and often advantageous in small and moderately sized networks, multi-beam configurations become progressively more beneficial, and eventually necessary, as network size and interference complexity increase.

\begin{table}[H]
\centering
\caption{Impact of antenna technology : sinble-beam versus multi-beam}
\label{tab:single_vs_multi}
\resizebox{\linewidth}{!}{%
\begin{tabular}{c c cc cc cc cc}
\toprule
Network size & Instance
 & \multicolumn{2}{c}{Overall} 
 & \multicolumn{2}{c}{Traffic Scenario $A$} 
 & \multicolumn{2}{c}{Traffic Scenario $B$} 
 & \multicolumn{2}{c}{Traffic Scenario $C$} \\
\cmidrule(lr){3-4}\cmidrule(lr){5-6}\cmidrule(lr){7-8}\cmidrule(lr){9-10}
 & 
& Single-Beam & Multi-Beam 
& Single-Beam & Multi-Beam 
& Single-Beam & Multi-Beam 
& Single-Beam & Multi-Beam \\

\midrule
10 &1  & 58.4576 & 45.8327 & 160.0000 & 120.1850 & 33.1407 & 27.5494 & 58.4232 & 45.6803 \\
   &2  & 58.4576 & 44.7354 & 160.0000 & 107.1110 & 33.1407 & 27.4191 & 58.4232 & 45.5965 \\
   &3  & 58.4576 & 38.4504 & 160.0000 & 54.4444  & 33.1407 & 34.3580 & 58.4232 & 38.4974 \\
   &4  & 58.4576 & 47.0703 & 160.0000 & 120.5190 & 33.1407 & 31.0148 & 58.4232 & 45.9169 \\
   &5  & 58.4576 & 53.8214 & 16.00000 & 107.5190 & 33.1407 & 32.8012 & 58.4232 & 57.6193 \\
   &6  &  58.4576& 44.1254 & 160.0000 & 120.4440 & 33.1407 & 25.1056 & 58.4232 & 44.0955 \\
   &7  & 58.4576 & 50.6161 & 160.0000 & 120.5190 & 33.1407 & 32.9000 & 58.4232 & 50.7364 \\
   &8  & 58.4576 & 41.5601 & 160.0000 & 54.2593  & 33.1407 & 30.6728 & 58.4232 & 45.4163 \\
   &9  & 58.4719 & 47.7473 & 60.9074 & 120.3700 & 53.8889 & 27.6481 & 60.4590 & 48.7190 \\
   &10 & 58.4576 & 41.7270 & 160.0000 & 55.2222  & 33.1407 & 30.6630 & 58.4232 & 45.5720 \\

\midrule
15 & 1  & 46.2707 & 36.2287 & 81.2143  & 54.2976  & 41.3095 & 30.8488 & 44.3833 & 36.6601 \\
   & 2  &  44.6879 & 35.2272 & 160.0000      & 107.4050 & 23.7673 & 19.7792 & 40.7342 & 33.9290 \\
   & 3  & 44.6879 & 34.2585 & 160.0000      & 54.6786  & 23.7673 & 31.0036 & 40.7342 & 33.3334 \\
   & 4  & 44.6879 & 33.0844 & 160.0000      & 54.3571  & 23.7673 & 27.6012 & 40.7342 & 33.1669 \\
   & 5  & 44.6879 & 33.0520 & 160.0000      & 41.1667  & 23.7673 & 30.8429 & 40.7342 & 33.1423 \\
   & 6  & 46.3375 & 40.7919 & 81.0952  & 41.1667  & 41.4048 & 40.7381 & 44.4591 & 40.7720 \\
   & 7  & 44.6879 & 36.3009 & 160.0000      & 41.3095  & 23.7673 & 34.1825 & 40.7342 & 36.7341 \\
   & 8  & 44.6879 & 34.1301 & 160.0000      & 54.2500  & 23.7673 & 30.9202 & 40.7342 & 33.2201 \\
   & 9  & 44.6879 & 37.9785 & 160.0000      & 107.4760 & 23.7673 & 20.7670 & 40.7342 & 37.8971 \\
   & 10 & 46.4886 & 38.9918 & 81.1429  & 61.0119  & 41.5714 & 27.4619 & 44.6154 & 42.0043 \\

\midrule
20 & 1 & 45.8989 & 33.3545 & 81.1579 & 60.8246 &41.4912  & 27.7939 &  43.6954& 32.7011 \\
   & 2 & 36.5940 & 31.7469 & 160.0000 & 54.5263 &16.8122  & 27.7807 &31.0591  & 30.8826 \\
   & 3 &34.3389  & 31.6918 & 81.1579 & 54.6491 & 28.3684 & 27.7032 & 31.4718 & 30.8165 \\
   & 4 & 32.8353 & 32.4574 & 81.2105 & 61.3070 &  24.2368& 27.9825 &31.0877  & 31.0887 \\
   & 5 & 34.2139 & 31.9851 &  81.2281& 54.7982 & 28.2175 & 28.0146 &31.3353  & 31.1186 \\
   & 6 & 36.5940 & 32.7510 &  160.0000& 41.1754 & 16.8122 & 30.9474 & 31.0591 & 32.5998 \\
   & 7 & 37.4280 & 31.6522 & 61.4649 & 54.3246 &33.3211  & 27.6974 & 36.4768 & 30.7956 \\
   & 8 &  34.2534 & 32.8614 & 81.2281 & 54.3158 &28.2632  & 27.8026 & 31.3767 & 32.7089 \\
   & 9 & 32.9055 & 31.7425 &81.2807  & 41.4035 & 24.3070 & 27.6579 &31.1579  & 32.5771 \\
   & 10 & 34.3314 & 33.9791 & 81.3684 & 54.4825 & 28.3333 & 31.1649 &31.4508  & 32.8233 \\

\midrule
30 & 1 &  27.5305& 25.1695 & 81.4138 & 54.6897 & 21.5057 & 21.1845 & 23.8075 & 23.4720 \\
   & 2 &  24.1305& 27.5827 &  81.2874& 61.2989 &17.6736  & 23.6839 & 20.2143 & 25.3176 \\
   & 3 &21.6817  & 22.2681 & 81.0345 & 61.1839 & 14.8976 & 17.3379 & 17.6546 & 19.8686 \\
   & 4 & 24.0397 & 25.8771 & 81.5517 & 61.1092 &13.5421  & 21.4239 & 22.0996 & 23.6997 \\
   & 5 &32.0735  & 24.7545 & 61.7529 & 54.7586 & 28.2931 & 21.2665 & 30.2537 & 22.7479 \\
   & 6 & 2.3452 & 25.2939 &2.8448 & 61.1379 &2.2900  & 21.3236 & 2.3103 & 22.7985 \\
   & 7 & 22.5855 & 22.3195 & 61.6207 & 61.4368 & 17.6420 & 17.3661 &20.1778  & 19.9065 \\
   & 8 &  2.3257 & 22.5007 & 3.0575 & 61.3103 &2.2452  & 17.9899 & 2.2744 & 19.9048 \\
   & 9 &  27.6073& 24.8529 & 81.2759 & 61.2874 & 21.6059 & 20.2057 &  23.8994& 22.6221 \\
   & 10 & 27.8671 & 25.0972 & 61.7241 & 54.7989 & 23.9531 & 21.0918 & 25.5919 & 23.3872 \\

\bottomrule
\end{tabular}%
}
\end{table}

\section{Sensitivity analysis of the objective function}
\label{sec:sensitivity_objective}

As detailed in Section \ref{sec:problem}, the objective function to be maximized is constructed as the combination of three distinct components, each corresponding to one of the scenarios $A$, $B$, and $C$. Within each component, performance is evaluated through a balance between a worst-case (minimum) measure and an average performance measure, with this trade-off explicitly governed by the parameter $p$. Furthermore, to reflect the relative importance of each scenario in the overall objective, a specific weight $\omega_X$ is assigned to every scenario $X \in \{A, B, C\}$, thereby allowing the formulation to prioritize certain scenarios over others.

Since the parameters $\omega_A$, $\omega_B$, $\omega_C$, and $p$ are, at least in part, determined by design considerations rather than being uniquely prescribed, a comprehensive sensitivity analysis is undertaken to examine how variations in these parameters affect the results. This analysis aims to verify whether the results remain consistent or instead depend critically on a specific choice of tuning.

\subsection{Sensitivity analysis with respect to the scenario weights}

The objective function can be rewritten in the following generic form:
$$\sum_{X\in\{A,B,C\} } \omega_X F_X
$$
where 
$$F_X=
\min_{[u,v] \in E} \frac{TP_{uv}}{n^X_{uv}}
+ 
p\;\operatorname*{mean}_{[u,v] \in E} \frac{TP_{uv}}{n^X_{uv}}.
$$
As explained in \cite{WH1}, for normalization purposes, the weights $\omega_A$, $\omega_B$, and $\omega_C$ are chosen so that they satisfy the following constraint, ensuring a consistent scaling across the different components of the objective function:
\[
\omega_A + \omega_B + \frac{\omega_C}{|V|-1} = 1.
\]

To examine the effect of the relative importance assigned to the three traffic scenarios, we consider several representative weight configurations $(\omega_A,\omega_B,\frac{\omega_C}{|V|-1})$, each capturing a different set of strategic priorities. The following configurations are evaluated.

\begin{itemize}[noitemsep]
    \item Baseline: $\left(\frac{1}{13}, \frac{4}{13}, \frac{8}{13}\right)$,
    \item Balanced: $\left(\frac{1}{3}, \frac{1}{3}, \frac{1}{3}\right)$,
    \item $A$-dominant: $\left(\frac{3}{5}, \frac{1}{5}, \frac{1}{5}\right)$,
    \item $B$-dominant: $\left(\frac{1}{5}, \frac{3}{5}, \frac{1}{5}\right)$,
    \item $C$-dominant: $\left(\frac{1}{5}, \frac{1}{5}, \frac{3}{5}\right)$,
    \item Moderately C-oriented (MCO): $\left(\frac{3}{20}, \frac{7}{20}, \frac{1}{2}\right)$,
    \item Strongly C-oriented (SCO): $\left(\frac{1}{20}, \frac{1}{4}, \frac{7}{10}\right)$.
\end{itemize}

The baseline configuration corresponds to the weight setting adopted in the main experiments. The balanced configuration assigns equal importance to all three scenarios. The dominant configurations place greater emphasis on a single scenario while maintaining moderate weights for the others. Finally, the last two configurations gradually increase the weight of scenario $C$, allowing us to assess how sensitive the solutions are to a progressive shift in its importance.

\begin{table}[H]
\centering
\scriptsize
\begin{tabular}{c|c|ccccccc}
\toprule
\textbf{$F_x$} & Network size 
& \textbf{Baseline} 
& \textbf{$A$-dominant} 
& \textbf{$B$-dominant} 
& \textbf{$C$-dominant} 
& \textbf{Balanced} 
& \textbf{MCO} 
& \textbf{SCO} \\
\hline

\multirow{4}{*}{$F_A$} 
& 10 & 98.06 & 107.22 & 101.23 & 103.28 & 107.22 & 100.66 & 82.33 \\
& 15 & 61.71 & 104.67 & 80.21  & 100.78 & 102.02 & 81.41  & 55.72 \\
& 20 & 74.28 & 107.35 & 60.50 & 68.48 & 83.72 & 55.19 & 49.51 \\
& 30 & 59.30 & 71.31 & 62.66 & 61.37 & 72.56 & 59.90 & 57.95 \\

\hline

\multirow{4}{*}{$F_B$} 
& 10 & 30.01 & 28.10 & 30.90 & 28.76 & 28.10 & 29.43 & 32.40 \\
& 15 & 29.31 & 18.52 & 26.13 & 20.41 & 19.24 & 27.43 & 29.77 \\
& 20 & 28.45 & 14.18 & 26.73 & 23.30 & 18.11 & 28.45 & 29.33 \\
& 30 & 20.19 & 14.19 & 18.25 & 19.21 & 13.91 & 18.66 & 20.99 \\

\hline

\multirow{4}{*}{$F_C$} 
& 10 & 46.79 & 43.97 & 44.40 & 46.29 & 43.97 & 46.61 & 47.41 \\
& 15 & 36.09 & 31.88 & 33.53 & 32.32 & 31.41 & 34.73 & 36.72 \\
& 20 & 31.81 & 23.34 & 30.19 & 29.58 & 25.35 & 31.20 & 31.73 \\
& 30 & 22.47 & 17.14 & 20.55 & 21.58 & 17.58 & 20.80 & 22.74 \\
\hline
\end{tabular}
\caption{Average $F_X$ values across scenario weight configurations (see Appendix for details)}
\label{tab:sensitivity_weights}
\end{table}

The average results reported in Table~\ref{tab:sensitivity_weights}, which are obtained by aggregating the detailed outcomes presented in Tables~\ref{tab:scenario_A}--\ref{tab:scenario_C} in the Appendix, offer a comprehensive and global perspective on how the weights $\omega_A$, $\omega_B$, and $\omega_C$ influence the resulting solutions across the different tested configurations.

More specifically, these aggregated results allow us to clearly observe the extent to which variations in the weighting scheme affect the performance levels associated with each scenario. In this context, changes in the average values of the scenario metrics $F_A$, $F_B$, and $F_C$ can be directly interpreted as meaningful indicators of the stability or variability of the underlying topology. Indeed, when the scenario values remain unchanged for a given instance despite modifications in the weights, this implies that the resulting topology is preserved and therefore structurally identical. Conversely, any observed variation in these values reflects a modification in the solution, indicating that the topology has changed in response to the altered weighting configuration.

For small instances (10 nodes), the average values remain relatively close across multiple configurations, indicating a generally high level of solution stability. For instance, Table~\ref{tab:sensitivity_weights} shows that the average value of $F_A$ increases from $98.06$ (Baseline) to $107.22$ (Balanced), which reflects only a moderate variation at the aggregate level.
However, a more detailed examination reveals that this average change conceals a mixed behaviour between stability and variability at the instance level. In particular, comparing the Baseline and Balanced configurations, $5$ out of $10$ instances (namely, instances 1, 2, 4, 6, and 7) preserve exactly the same values of $F_A$, $F_B$, and $F_C$, indicating unchanged topologies, while the remaining $5$ instances exhibit variations in their scenario values, which correspond to changes in the underlying solutions.
For example, the first instance in Table~\ref{tab:scenario_A} maintains identical values $F_A=120.19$, $F_B=27.55$, and $F_C=45.68$ across both configurations, thereby reflecting full stability of the topology. In contrast, the third  instance exhibits noticeable changes in these values, which directly indicates a modification of the associated topology.

For medium-sized instances (15 and 20 nodes), the differences in average values become more marked, especially for scenario~$A$. For instance, for 20 nodes, $F_A$ rises substantially from $74.28$ in the Baseline configuration to $107.35$ in the $A$-dominant configuration. This pronounced change in the aggregated results reflects modifications in the selected topology for a significant share of instances, indicating that the optimization process systematically adapts the network structure to better accommodate the emphasized scenario. By contrast, the variations observed for $F_B$ and $F_C$ are more moderate, suggesting that these scenarios have a comparatively weaker influence on inducing topology changes.

For large instances (30 nodes), the differences become more pronounced, indicating greater instability of the solutions compared to the Baseline configuration. For example, the average $F_C$ value increases from $22.47$ (Baseline) to $22.74$ (Strongly C-oriented). Although this change appears modest at the aggregate level, it in fact masks a substantial number of topology modifications: the detailed tables show that $9$ out of $10$ instances exhibit different values of $F_C$, confirming that the underlying solutions are frequently altered.

More generally, the observed differences between the Baseline configuration and the strongly biased configurations indicate that the optimal solutions become increasingly sensitive to the choice of weights. This heightened sensitivity can be attributed to the larger size of the network, which expands the solution space and enables the algorithm to select a wider variety of topologies depending on the imposed priorities.

Overall, the joint analysis of the average values and the detailed instance-level results shows that the weights $\omega_A$, $\omega_B$, and $\omega_C$ have a direct impact on topology stability. In particular, scenario $A$, due to its higher magnitude, triggers the most pronounced changes in topology, whereas scenarios $B$ and $C$ lead to more gradual and limited effects. Consequently, the weighting scheme acts not only as an evaluation criterion but also as a structural mechanism that steers the selection of network topologies.

\subsection{Sensitivity analysis of the worst-case vs average performance trade-off}

As mentioned in the previous section, each component $F_X$ of the objective function is constructed as a trade-off between worst-case and average performance. More precisely, this can be written in the form $F_X = F_X^{\min} + pF_X^{\text{mean}}$, where 
$$\displaystyle F_X^{\min}=\min_{[u,v] \in E} \frac{TP_{uv}}{n^X_{uv}}\qquad\text{ and }\qquad F_X^{\text{mean}}=\operatorname*{mean}_{[u,v] \in E} \frac{TP_{uv}}{n^X_{uv}}$$

The first term $F_X^{\min}$ captures the worst-case (minimum) behavior of the system, while the second term $F_X^{\text{mean}}$ accounts for its average performance. The parameter $p$ is used to control the relative importance assigned to the average component with respect to the worst-case one, thereby continuously tuning the balance between conservative and performance-oriented objectives.

To analyze the impact of the trade-off between worst-case and average performance, we introduce the following measure:
\[
F_X(\lambda) = (2-\lambda)F_X^{\min}+ \lambda pF_X^{\text{mean}},
\]

\noindent where $\lambda \in [0,2]$ controls the balance between the two terms. By setting $\lambda = 1$, we recover the original value of $F_X$, i.e., $F_X(1) = F(X)$. To further explore the sensitivity of the model, we also consider additional parameter settings beyond the standard range, namely $\lambda = 0$, $0.5$, $1.5$, and $2$.
These settings are used to examine different degrees of emphasis on the trade-off between worst-case and average performance. In particular, the extreme cases $\lambda = 0$ and $\lambda = 2$ correspond, respectively, to a purely worst-case objective and to a strongly average-dominated objective.

Instead of minimizing $\sum_{X\in\{A,B,C\} } \omega_X F_X$, we minimize $\sum_{X\in\{A,B,C\} } \omega_X F_X(\lambda)$, using the aforementioned values of $\lambda$. 
Results are reported in Table~\ref{tab:lambda_final}, which summarizes the average values obtained from the detailed results provided in Tables~\ref{tab:lambda_scenario_A}--\ref{tab:lambda_scenario_C} in the Appendix. The analysis provides a clearer global understanding of the influence of the parameter $\lambda$ across different network sizes and scenarios. Overall, increasing $\lambda$ progressively shifts the optimization from worst-case–oriented solutions toward configurations that favor higher average throughput. However, the strength of this effect varies significantly depending on both the scenario considered and the size of the graph.

\begin{table}[H]
\centering
\scriptsize
\begin{tabular}{c|ccccc|ccccc}
\hline
\textbf{network size} & 
\textbf{Baseline} & $\lambda=0$ & $\lambda=0.5$ & $\lambda=1.5$ & $\lambda=2$
& \textbf{Baseline} & $\lambda=0$ & $\lambda=0.5$ & $\lambda=1.5$ & $\lambda=2$ \\
\hline
& \multicolumn{5}{c|}{$F_A^{\min}$} 
& \multicolumn{5}{c}{$F_A^{\mathrm{mean}}$} \\
\hline

10 & 94.90 & 79.30 & 94.90 & 94.90 & 39.00 & 123.61 & 113.60 & 123.61 & 123.61 & 108.99 \\
15 & 59.15 & 59.15 & 52.00 & 53.25 & 27.30 & 99.91  & 95.18  & 94.81  & 99.91  & 105.81 \\
20 & 50.70 & 53.95 & 50.05 & 56.05 & 26.00 & 96.75  & 93.26  & 96.61  & 95.48  & 109.91 \\
30 & 56.75 & 50.70 & 57.20 & 58.05 & 16.90 & 107.30 & 90.26  & 103.95 & 107.33 & 105.53 \\
\hline

& \multicolumn{5}{c|}{$F_B^{\min}$} 
& \multicolumn{5}{c}{$F_B^{\mathrm{mean}}$} \\
\hline

10 & 28.41 & 29.01 & 28.41 & 28.41 & 14.30 & 62.72 & 60.14 & 62.72 & 62.72 & 84.41 \\
15 & 27.26 & 27.75 & 29.36 & 28.93 & 11.09 & 60.74 & 57.49 & 61.83 & 62.03 & 84.60 \\
20 & 26.98 & 25.73 & 26.93 & 26.28 & 7.76  & 70.98 & 63.35 & 68.83 & 73.66 & 87.03 \\
30 & 18.52 & 17.12 & 18.38 & 17.67 & 2.52  & 68.57 & 57.32 & 67.58 & 69.49 & 84.46 \\
\hline

& \multicolumn{5}{c|}{$F_C^{\min}$} 
& \multicolumn{5}{c}{$F_C^{\mathrm{mean}}$} \\
\hline

10 & 44.93 & 44.93 & 44.93 & 44.93 & 18.59 & 8.03 & 7.98 & 8.03 & 8.03 & 9.73 \\
15 & 34.93 & 33.67 & 34.74 & 34.78 & 12.97 & 4.60 & 4.36 & 4.63 & 4.72 & 6.16 \\
20 & 28.80 & 28.13 & 28.65 & 28.80 & 8.98  & 3.79 & 3.43 & 3.71 & 3.76 & 4.65 \\
30 & 20.57 & 19.48 & 20.30 & 20.07 & 4.02  & 2.43 & 2.02 & 2.39 & 2.45 & 2.94 \\

\hline

\end{tabular}
\caption{Average $F_X^{\min}$ and $ F_X^{\text{mean}}$ values for different $\lambda$ settings (see Appendix for details)}
\label{tab:lambda_final}
\end{table}

For Scenario~$A$, the averaged results indicate a relatively low sensitivity to $\lambda$, particularly for small and medium-sized networks. For instance, for 10 nodes, the average values of $F_A^{\min}$ and $ F_A^{\text{mean}}$ remain unchanged for $\lambda = 0.5$, $1$ (baseline), and $1.5$, suggesting that many instances retain the same optimal topology across these settings. This is consistent with the detailed tables, which show that a majority of instances produce identical solutions for these intermediate values of $\lambda$.
A similar pattern of stability is observed for 15 and 20 nodes, although more noticeable variations begin to emerge as the network size increases. For example, for 20 nodes, the average minimum $F_A^{\min}$ value increases from $50.70$ to $56.05$ at $\lambda = 1.5$, indicating that a subset of instances benefits from a more balanced trade-off between minimum and mean objectives.
For larger networks with 30 nodes, the impact of $\lambda$ becomes more pronounced. While the average mean $F_A^{\text{mean}}$ remains relatively stable, the average minimum $F_A^{\min}$ decreases sharply for $\lambda = 2$, dropping from $56.75$ to $16.90$. The detailed tables indicate that this reduction is driven by several instances in which the minimum value falls to zero. Hence, although Scenario~$A$ appears globally stable at first glance, the averaged results reveal a clear degradation when extreme values of $\lambda$ are used in large networks.

For Scenario~$B$, the averaged results reveal a substantially stronger sensitivity to $\lambda$. As $\lambda$ increases, the average performance $F_B^{\text{mean}}$ improves markedly across all network sizes. For instance, with 20 nodes, it rises from $70.98$ to $87.03$ at $\lambda = 2$, and a comparable trend is observed for 30 nodes. The detailed tables further confirm that this improvement is consistent across a large majority of instances. However, this improvement comes at the cost of a substantial deterioration in the minimum performance. On average, $F_B^{\min}$ decreases sharply, for example from $26.98$ to $7.76$ with 20 nodes, and from $18.52$ to $2.52$ with 30 nodes. The detailed results show that this decline is not marginal: a large proportion of instances, often more than half, exhibit very low or even zero minimum values when $\lambda = 2$. This clearly indicates that, in Scenario~$B$, the optimization increasingly sacrifices fairness in favor of global efficiency, particularly as the network size grows.

For Scenario~$C$, a similar trend is observed, with an even stronger sensitivity emerging for larger networks. The average performance $F_C^{\text{mean}}$ increases progressively with $\lambda$, for instance from $2.43$ to $2.94$ with 30 nodes. The detailed tables confirm that this improvement is shared by most instances.
In contrast, the average minimum performance $F_C^{\min}$ decreases substantially, falling from $20.57$ to $4.02$ for 30 nodes. The detailed results further indicate that several instances reach zero or near-zero values when $\lambda = 2$. For smaller network sizes with 10 and 15 nodes, the decrease is noticeably less pronounced, with variations remaining moderate and much more limited in magnitude compared to larger networks, highlighting that the effect of $\lambda$ becomes more pronounced as the graph size increases.

Overall, combining the average results with the detailed instance-level analysis provides a clear interpretation of the role of $\lambda$. Intermediate values such as $\lambda = 0.5$, $1$, and $1.5$ generally achieve a satisfactory balance between average performance and worst-case guarantees. In contrast, the extreme case $\lambda = 2$ systematically leads to a loss of stability, particularly in large networks, where a significant number of instances exhibit very low minimum throughput. From a topological standpoint, these findings indicate that solutions obtained with larger values of $\lambda$ tend to concentrate performance on a subset of links. This improves average performance but comes at the expense of overall fairness.

\section{Conclusion}\label{sec:conclusion}

This comprehensive sensitivity analysis offers a broad perspective on how structural, technological, and modeling parameters impact the performance of tactical wireless networks.

The analysis of master hub selection confirms that both hub centrality and connectivity must be considered jointly. Strategies that combine degree and eccentricity consistently yield stable and high-quality solutions, whereas peripheral choices such as the \texttt{LeafOnly} strategy systematically degrade performance. The baseline strategy provides a good compromise between computational efficiency and stability of the objective value, highlighting the importance of selecting hubs that are both structurally central and well connected.

The analysis of the PMP successor limit reveals a clear threshold effect. For small instances, the constraint is effectively inactive and has no impact on solution quality. For medium-sized networks, too restrictive values lead to a noticeable degradation in performance, whereas values above a moderate level do not yield further improvements. For larger instances, the influence of this parameter becomes negligible. Overall, these results confirm that the baseline PMP configuration is structurally adequate and prevents unnecessary enlargement of the model.

Antenna beam width shows a pronounced scale-dependent effect. In small networks, directivity has a limited impact because interference levels remain low. As the network size increases, insufficient beam refinement results in statistically significant performance degradation. However, beyond a certain level of directivity, the benefits saturate and diminishing returns are observed, revealing a technological trade-off between interference mitigation and connectivity flexibility.

The comparison between single-beam and multi-beam technologies highlights a structural transition that depends on network size. In small and moderately sized networks, single-beam antennas remain competitive and are often even superior, since interference levels are still manageable. However, as the network grows in density and complexity, multi-beam configurations progressively outperform single-beam systems by enabling simultaneous directional transmissions and mitigating congestion effects. For large-scale instances, multi-beam technology becomes essential to maintain throughput and avoid a structural degradation of performance.


The sensitivity analysis of the objective function confirms that the proposed formulation is sound and consistent.  Variations in the scenario weights $\omega_A$, $\omega_B$, and $\omega_C$ adjust the relative importance assigned to different traffic patterns. More precisely, the impact of these weights intensifies as the network size grows. For small instances, solutions remain relatively stable across weight configurations, whereas larger networks exhibit increased sensitivity to the weighting scheme.

Furthermore, adjusting the balance between the minimum and mean components of the objective function reveals a clear trade-off between minimum throughput and average performance. A smooth transition is observed from a worst-case-oriented objective to a balanced compromise, and ultimately to a strongly average-oriented objective, where overall network performance improves at the expense of fairness and minimum throughput, particularly in larger networks.


Overall, the results show that the proposed framework preserves structural consistency under moderate parameter variations, while also highlighting clear scale-dependent technological transitions. The topology design remains coherent across different modeling settings, and the observed performance variations are mainly governed by identifiable threshold effects rather than abrupt or unpredictable structural changes.

These findings indicate that future research could further explore parameter-driven strategies for topology shaping, by systematically varying structural and technological parameters to guide the optimization process toward targeted performance profiles. In particular, investigating controlled parameter tuning aimed at enhancing specific performance metrics while enforcing fixed operational constraints appears to be a promising avenue for refining topology design in tactical wireless networks.

\bibliographystyle{acm} 
\bibliography{REF}

@mastersthesis{2,
  title={Tactical wireless network design for challenging environments},
  author={Perreault, Vincent},
  year={2022},
  school={Polytechnique Montr{\'e}al}
}

@article{WH1,
  title={Tabu Search for Tactical Wireless Network Design in
Challenging Environments},
  author={Ahmed Zaid, Wissem and Hertz, Alain},
  year={2026},
  journal={arXiv preprint arXiv:2604.18318}
}

@inproceedings{mumey,
  title={Topology control in multihop wireless networks with multi-beam smart antennas},
  author={Mumey, Brendan and Judson, Ivan and Tang, Jian and Xing, Yun},
  booktitle={2012 International Conference On Computing, Networking and Communications (ICNC)},
  pages={1020--1024},
  year={2012},
  organization={IEEE}
}

@inproceedings{zhou,
  title={On capacity optimization in multi-radio multi-channel wireless networks with directional antennas},
  author={Zhou, Lei and Cao, Xianghui and Liu, Lu and Cai, Lin and Tian, Xiaohua and Cheng, Yu},
  booktitle={2015 IEEE International Conference on Communications (ICC)},
  pages={3745--3750},
  year={2015},
  organization={IEEE}
}

@inproceedings{hamami,
  title={Joint resource allocation in multi-radio multi-channel wireless mesh networks with practical sectored antennas},
  author={Hamami, Nasrin Sadeghianpour and Chuah, Teong Chee and Tan, Su Wei},
  booktitle={2010 International Conference on Computer Applications and Industrial Electronics},
  pages={316--321},
  year={2010},
  organization={IEEE}
}

@article{wang2016,
  title={Infrastructure communication sensitivity analysis of wireless sensor networks},
  author={Wang, Chaonan and Xing, Liudong and Vokkarane, Vinod M and Sun, Yan},
  journal={Quality and Reliability Engineering International},
  volume={32},
  number={2},
  pages={581--594},
  year={2016},
  publisher={Wiley Online Library}
}

@inproceedings{khuller1994low,
  title={Low degree spanning trees of small weight},
  author={Khuller, Samir and Raghavachari, Balaji and Young, Neal},
  booktitle={Proceedings of the twenty-sixth annual ACM symposium on Theory of computing},
  pages={412--421},
  year={1994}
}

@article{shi2018modeling,
  title={Modeling and analysis of point-to-multipoint millimeter wave backhaul networks},
  author={Shi, Jia and Lv, Lu and Ni, Qiang and Pervaiz, Haris and Paoloni, Claudio},
  journal={IEEE Transactions on Wireless Communications},
  volume={18},
  number={1},
  pages={268--285},
  year={2018},
  publisher={IEEE}
}

@inproceedings{kumar2006topology,
  title={A topology control approach to using directional antennas in wireless mesh networks},
  author={Kumar, Umesh and Gupta, Himanshu and Das, Samir R},
  booktitle={2006 IEEE International Conference on Communications},
  volume={9},
  pages={4083--4088},
  year={2006},
  organization={IEEE}
}

@inproceedings{promponas2023optimizing,
  title={Optimizing sectorized wireless networks: Model, analysis, and algorithm},
  author={Promponas, Panagiotis and Chen, Tingjun and Tassiulas, Leandros},
  booktitle={Proceedings of the Twenty-fourth International Symposium on Theory, Algorithmic Foundations, and Protocol Design for Mobile Networks and Mobile Computing},
  pages={141--150},
  year={2023}
}

@article{demvsar2006,
  title={Statistical comparisons of classifiers over multiple data sets},
  author={Dem{\v{s}}ar, Janez},
  journal={Journal of Machine learning research},
  volume={7},
  number={Jan},
  pages={1--30},
  year={2006}
}

@incollection{wilcoxon1992,
  title={Individual comparisons by ranking methods},
  author={Wilcoxon, Frank},
  booktitle={Breakthroughs in statistics: Methodology and distribution},
  pages={196--202},
  year={1992},
  publisher={Springer}
}

@article{friedman1937,
  title={The use of ranks to avoid the assumption of normality implicit in the analysis of variance},
  author={Friedman, Milton},
  journal={Journal of the american statistical association},
  volume={32},
  number={200},
  pages={675--701},
  year={1937},
  publisher={Taylor \& Francis}
}

@article{marina,
  title={A topology control approach for utilizing multiple channels in multi-radio wireless mesh networks},
  author={Marina, Mahesh K and Das, Samir R and Subramanian, Anand Prabhu},
  journal={Computer networks},
  volume={54},
  number={2},
  pages={241--256},
  year={2010},
  publisher={Elsevier}
}

@inproceedings{zhou2008,
  title={Practical routing and channel assignment scheme for mesh networks with directional antennas},
  author={Zhou, Wei and Chen, Xi and Qiao, Daji},
  booktitle={2008 IEEE International Conference on Communications},
  pages={3181--3187},
  year={2008},
  organization={IEEE}
}

@inproceedings{8,
  title={Design and optimization of a tiered wireless access network},
  author={Son, In Keun and Mao, Shiwen},
  booktitle={2010 Proceedings IEEE INFOCOM},
  pages={1--9},
  year={2010},
  organization={IEEE}
}

@inproceedings{10,
  title={Topology design of hierarchical hybrid fiber-vdsl access networks with aco},
  author={Zhao, Rong and Liu, Hanjie and Lehnert, Ralf},
  booktitle={2008 Fourth Advanced International Conference on Telecommunications},
  pages={232--237},
  year={2008},
  organization={IEEE}
}

@article{12,
  title={Cost-optimal topology planning of hierarchical access networks},
  author={G{\'o}dor, Istv{\'a}n and Magyar, G{\'a}bor},
  journal={Computers \& operations research},
  volume={32},
  number={1},
  pages={59--86},
  year={2005},
  publisher={Elsevier}
}

@inproceedings{13,
  title={Minimum cost wireless broadband overlay network planning},
  author={Lin, Peng and Ngo, Hung and Qiao, ChunMing and Wang, Xin and Wang, Ting and Qian, DaYou},
  booktitle={2006 International Symposium on a World of Wireless, Mobile and Multimedia Networks (WoWMoM'06)},
  pages={7--pp},
  year={2006},
  organization={IEEE}
}

@article{15,
  title={Joint design of hierarchical topology control and routing design for heterogeneous wireless sensor networks},
  author={An, Jian and Qi, Ling and Gui, Xiaolin and Peng, Zhenlong},
  journal={Computer Standards \& Interfaces},
  volume={51},
  pages={63--70},
  year={2017},
  publisher={Elsevier}
}

@article{shukla2023angle,
  title={Angle based critical nodes detection (ABCND) for reliable industrial wireless sensor networks},
  author={Shukla, Shailendra},
  journal={Wireless Personal Communications},
  volume={130},
  number={2},
  pages={757--775},
  year={2023},
  publisher={Springer}
}

@article{wzorek2021router,
  title={Router and gateway node placement in wireless mesh networks for emergency rescue scenarios},
  author={Wzorek, Mariusz and Berger, Cyrille and Doherty, Patrick},
  journal={Autonomous intelligent systems},
  volume={1},
  number={1},
  pages={14},
  year={2021},
  publisher={Springer}
}

\newpage
\textbf{Appendix}

\begin{table}[H]
\centering
\caption{$F_A$ values for different scenario weights.}
\label{tab:scenario_A}
\resizebox{\linewidth}{!}{%
\begin{tabular}{c c | ccccccc}
\toprule
Network size & Instance & \multicolumn{7}{c}{Scenario $A$} \\
\cmidrule(lr){3-9}

& 
& Base & $A$-dominant & $B$-dominant & $C$-dominant & Balanced & MCO & SCO \\

\midrule

\multirow{10}{*}{10}
& 1  & 120.19 & 120.19 & 120.19 & 120.19 & 120.19 & 120.19 & 55.04 \\
& 2  & 107.11 & 107.11 & 107.11 & 107.11 & 107.11 & 107.11 & 54.02 \\
& 3  & 54.44  & 80.63  & 80.63  & 80.63  & 80.63  & 54.44  & 54.44 \\
& 4  & 120.52 & 120.52 & 120.52 & 120.52 & 120.52 & 120.52 & 120.52 \\
& 5  & 107.52 & 120.52 & 120.52 & 107.52 & 120.52 & 107.52 & 107.52 \\
& 6  & 120.44 & 120.44 & 120.44 & 120.44 & 120.44 & 120.44 & 81.37 \\
& 7  & 120.52 & 120.52 & 120.52 & 120.52 & 120.52 & 120.52 & 120.52 \\
& 8  & 54.26  & 80.67  & 40.94  & 54.26  & 80.67  & 54.26  & 54.26 \\
& 9  & 120.37 & 120.41 & 120.41 & 120.37 & 120.41 & 120.37 & 120.37 \\
& 10 & 55.22  & 81.19  & 61.00  & 81.19  & 81.19  & 81.19  & 55.22 \\

\midrule
\multirow{10}{*}{15}
& 1  & 54.30  & 81.38  & 61.27  & 81.38  & 81.38  & 54.54  & 54.30 \\
& 2  & 107.41 & 120.36 & 120.41 & 120.43 & 120.31 & 120.36 & 42.07 \\
& 3  & 54.68  & 120.36 & 107.48 & 107.41 & 120.31 & 107.33 & 54.68 \\
& 4  & 54.36  & 107.24 & 54.36  & 81.17  & 81.00  & 60.79  & 60.79 \\
& 5  & 41.17  & 80.93  & 80.48  & 80.83  & 81.14  & 81.17  & 41.17 \\
& 6  & 41.17  & 107.60 & 41.36  & 107.60 & 107.38 & 41.17  & 41.17 \\
& 7  & 41.31  & 120.24 & 107.26 & 120.24 & 120.24 & 120.29 & 41.31 \\
& 8  & 54.25  & 107.38 & 67.81  & 107.38 & 107.45 & 67.81  & 54.25 \\
& 9  & 107.48 & 120.29 & 107.41 & 120.45 & 120.29 & 107.38 & 107.48 \\
& 10 & 61.01  & 80.93  & 54.24  & 80.93  & 80.69  & 54.24  & 61.01 \\

\midrule
\multirow{10}{*}{20}
& 1  & 60.82 & 107.39 & 54.68 & 61.32 & 81.25 & 54.68 & 41.16 \\
& 2  & 54.53 & 120.53 & 61.10 & 61.25 & 81.19 & 54.48 & 61.21 \\
& 3  & 54.65 & 81.07  & 54.52 & 54.42 & 60.98 & 54.23 & 53.96 \\
& 4  & 61.31 & 107.42 & 81.11 & 107.23 & 80.98 & 61.15 & 41.23 \\
& 5  & 54.80 & 107.28 & 54.83 & 61.33 & 81.21 & 54.80 & 41.08 \\
& 6  & 41.18 & 107.21 & 60.79 & 107.33 & 107.28 & 54.39 & 41.38 \\
& 7  & 54.32 & 81.35  & 54.43 & 54.65 & 81.32 & 41.46 & 41.17 \\
& 8  & 120.35 & 120.35 & 68.00 & 61.25 & 61.22 & 61.18 & 61.09 \\
& 9  & 120.44 & 120.44 & 54.52 & 54.54 & 81.33 & 54.46 & 54.48 \\
& 10 & 120.44 & 120.44 & 60.98 & 61.47 & 120.40 & 61.04 & 54.31 \\

\midrule
\multirow{10}{*}{30}
& 1  & 54.69 & 81.29 & 61.15 & 61.42 & 61.38 & 61.16 & 54.61 \\
& 2  & 61.30 & 81.29 & 68.09 & 61.40 & 107.53 & 61.32 & 61.43 \\
& 3  & 61.18 & 67.98 & 61.04 & 54.43 & 67.98 & 54.43 & 54.53 \\
& 4  & 61.11 & 67.94 & 61.39 & 61.46 & 67.98 & 61.45 & 61.44 \\
& 5  & 54.76 & 68.10 & 61.39 & 61.36 & 61.36 & 61.07 & 54.43 \\
& 6  & 61.14 & 67.97 & 61.21 & 61.45 & 80.98 & 61.19 & 61.20 \\
& 7  & 61.44 & 68.14 & 61.51 & 68.16 & 67.89 & 54.51 & 61.28 \\
& 8  & 61.31 & 61.48 & 61.31 & 61.31 & 61.31 & 61.31 & 54.66 \\
& 9  & 61.29 & 67.97 & 61.52 & 61.26 & 67.94 & 61.21 & 61.29 \\
& 10 & 54.80 & 80.97 & 67.99 & 61.47 & 81.24 & 61.32 & 54.59 \\
\bottomrule
\end{tabular}}
\end{table}

\begin{table}[H]
\centering
\caption{$F_B$ values for different scenario weights.}
\label{tab:scenario_B}
\resizebox{\linewidth}{!}{%
\begin{tabular}{c c | ccccccc}
\toprule
Network size & Instance & \multicolumn{7}{c}{Scenario $B$} \\
\cmidrule(lr){3-9}

& 
& Base & $A$-dominant & $B$-dominant & $C$-dominant & Balanced & MCO & SCO \\
\midrule
\multirow{10}{*}{10}
& 1  & 27.55 & 27.55 & 27.55 & 27.55 & 27.55 & 27.55 & 32.53 \\
& 2  & 27.42 & 27.42 & 27.42 & 27.42 & 27.42 & 27.42 & 40.57 \\
& 3  & 34.36 & 27.70 & 27.70 & 27.70 & 27.70 & 34.36 & 34.36 \\
& 4  & 31.01 & 31.01 & 31.01 & 31.01 & 31.01 & 31.01 & 31.01 \\
& 5  & 32.80 & 32.83 & 32.83 & 32.80 & 32.83 & 32.80 & 32.80 \\
& 6  & 25.11 & 25.11 & 25.11 & 25.11 & 25.11 & 25.11 & 30.89 \\
& 7  & 32.90 & 32.90 & 32.90 & 32.90 & 32.90 & 32.90 & 32.90 \\
& 8  & 30.67 & 20.76 & 40.17 & 30.67 & 20.76 & 30.67 & 30.67 \\
& 9  & 27.65 & 30.89 & 30.89 & 27.65 & 30.89 & 27.65 & 27.65 \\
& 10 & 30.66 & 24.84 & 34.43 & 24.83 & 24.84 & 24.84 & 30.66 \\

\midrule
\multirow{10}{*}{15}
& 1  & 30.85 & 20.68 & 31.06 & 20.68 & 20.68 & 30.81 & 30.85 \\
& 2  & 19.78 & 19.76 & 18.00 & 19.82 & 20.68 & 19.76 & 23.40 \\
& 3  & 31.00 & 14.16 & 18.61 & 19.83 & 17.90 & 19.75 & 31.00 \\
& 4  & 27.60 & 17.91 & 27.60 & 20.60 & 20.61 & 27.55 & 27.55 \\
& 5  & 30.84 & 19.89 & 20.85 & 19.77 & 19.60 & 19.68 & 30.84 \\
& 6  & 40.74 & 19.83 & 40.69 & 19.83 & 20.75 & 40.74 & 40.74 \\
& 7  & 34.18 & 18.46 & 19.71 & 18.46 & 18.46 & 19.77 & 34.18 \\
& 8  & 30.92 & 18.59 & 27.72 & 18.59 & 17.97 & 27.72 & 30.92 \\
& 9  & 20.77 & 15.66 & 22.86 & 19.85 & 15.66 & 34.24 & 20.77 \\
& 10 & 27.46 & 21.21 & 34.24 & 27.65 & 20.10 & 34.24 & 27.46 \\

\midrule
\multirow{10}{*}{20}
& 1  & 27.79 & 15.52 & 31.09 & 27.84 & 15.09 & 31.13 & 30.71 \\
& 2  & 27.78 & 15.49 & 27.66 & 21.23 & 20.10 & 27.97 & 28.00 \\
& 3  & 27.70 & 9.58  & 27.75 & 25.17 & 27.83 & 27.60 & 30.69 \\
& 4  & 27.98 & 14.00 & 16.97 & 15.42 & 15.94 & 27.82 & 27.71 \\
& 5  & 28.01 & 14.04 & 28.03 & 25.35 & 15.83 & 28.01 & 27.53 \\
& 6  & 30.95 & 14.01 & 27.86 & 14.07 & 13.95 & 20.42 & 30.95 \\
& 7  & 27.70 & 13.99 & 27.74 & 23.43 & 16.48 & 31.13 & 27.51 \\
& 8  & 27.80 & 15.42 & 21.19 & 27.70 & 27.91 & 31.38 & 31.16 \\
& 9  & 27.66 & 15.66 & 27.85 & 24.89 & 13.97 & 27.79 & 27.93 \\
& 10 & 31.16 & 14.05 & 31.17 & 27.88 & 14.02 & 31.22 & 31.11 \\

\midrule
\multirow{10}{*}{30}
& 1  & 21.18 & 10.91 & 21.24 & 21.33 & 14.48 & 21.31 & 21.36 \\
& 2  & 23.68 & 11.49 & 14.63 & 20.54 & 9.48  & 23.63 & 21.64 \\
& 3  & 17.34 & 14.53 & 16.14 & 21.25 & 12.66 & 17.22 & 21.27 \\
& 4  & 21.42 & 18.04 & 21.37 & 14.40 & 17.96 & 16.10 & 21.53 \\
& 5  & 21.27 & 15.94 & 21.27 & 21.38 & 21.38 & 21.27 & 21.24 \\
& 6  & 21.32 & 11.68 & 15.66 & 16.51 & 9.20  & 12.82 & 21.41 \\
& 7  & 17.37 & 14.58 & 16.44 & 16.15 & 12.32 & 20.93 & 17.94 \\
& 8  & 17.99 & 17.31 & 17.99 & 17.99 & 17.99 & 17.99 & 21.07 \\
& 9  & 20.21 & 12.98 & 16.53 & 21.26 & 12.31 & 17.27 & 20.21 \\
& 10 & 21.09 & 14.43 & 21.25 & 21.31 & 11.30 & 18.01 & 21.19 \\
\bottomrule
\end{tabular}}
\end{table}

\begin{table}[H]
\centering
\caption{$F_C$ values for different scenario weights.}
\label{tab:scenario_C}
\resizebox{\linewidth}{!}{%
\begin{tabular}{c c | ccccccc}
\toprule
Network size & Instance & \multicolumn{7}{c}{Scenario $C$} \\
\cmidrule(lr){3-9}

& 
& Base & $A$-dominant & $B$-dominant & $C$-dominant & Balanced & MCO & SCO \\
\midrule

\multirow{10}{*}{10}
& 1  & 45.68 & 45.68 & 45.68 & 45.68 & 45.68 & 45.68 & 50.35 \\
& 2  & 45.60 & 45.60 & 45.60 & 45.60 & 45.60 & 45.60 & 45.50 \\
& 3  & 38.50 & 35.26 & 35.26 & 35.26 & 35.26 & 38.50 & 38.50 \\
& 4  & 45.92 & 45.92 & 45.92 & 45.92 & 45.92 & 45.92 & 45.92 \\
& 5  & 57.62 & 52.07 & 52.07 & 57.62 & 52.07 & 57.62 & 57.62 \\
& 6  & 44.10 & 44.10 & 44.10 & 44.10 & 44.10 & 44.10 & 45.79 \\
& 7  & 50.74 & 50.74 & 50.74 & 50.74 & 50.74 & 50.74 & 50.74 \\
& 8  & 45.42 & 30.74 & 40.28 & 45.42 & 30.74 & 45.42 & 45.42 \\
& 9  & 48.72 & 45.79 & 45.79 & 48.72 & 45.79 & 48.72 & 48.72 \\
& 10 & 45.57 & 43.83 & 38.56 & 43.83 & 43.83 & 43.83 & 45.57 \\

\midrule
\multirow{10}{*}{15}
& 1  & 36.66 & 31.74 & 33.39 & 31.74 & 31.74 & 36.65 & 36.66 \\
& 2  & 33.93 & 33.92 & 30.75 & 31.79 & 31.74 & 33.92 & 40.32 \\
& 3  & 33.33 & 33.92 & 28.46 & 33.98 & 30.65 & 33.90 & 33.33 \\
& 4  & 33.17 & 30.52 & 33.17 & 31.65 & 26.12 & 33.12 & 33.12 \\
& 5  & 33.14 & 31.81 & 26.26 & 31.70 & 31.58 & 30.44 & 33.14 \\
& 6  & 40.77 & 33.99 & 40.75 & 33.99 & 31.80 & 40.77 & 40.77 \\
& 7  & 36.73 & 33.86 & 33.85 & 33.86 & 33.86 & 31.75 & 36.73 \\
& 8  & 33.22 & 28.44 & 35.12 & 28.44 & 30.72 & 32.15 & 33.22 \\
& 9  & 37.90 & 33.94 & 36.78 & 34.00 & 33.94 & 37.82 & 37.90 \\
& 10 & 42.00 & 26.63 & 36.78 & 32.09 & 31.98 & 36.78 & 42.00 \\

\midrule
\multirow{10}{*}{20}
& 1  & 32.70 & 24.41 & 32.77 & 30.96 & 24.24 & 32.80 & 32.38 \\
& 2  & 30.88 & 26.19 & 30.77 & 31.21 & 24.79 & 29.44 & 31.10 \\
& 3  & 30.82 & 16.04 & 30.85 & 30.87 & 30.93 & 30.69 & 32.35 \\
& 4  & 31.09 & 24.34 & 21.25 & 24.30 & 24.58 & 30.93 & 32.62 \\
& 5  & 31.12 & 22.92 & 31.14 & 31.06 & 23.24 & 31.12 & 30.62 \\
& 6  & 32.60 & 21.75 & 30.94 & 23.70 & 23.58 & 29.02 & 32.61 \\
& 7  & 30.80 & 23.62 & 29.23 & 31.22 & 24.59 & 32.79 & 30.61 \\
& 8  & 32.71 & 26.10 & 29.33 & 30.82 & 31.02 & 33.04 & 32.82 \\
& 9  & 32.58 & 24.37 & 32.83 & 30.61 & 22.87 & 29.27 & 29.40 \\
& 10 & 32.82 & 23.68 & 32.83 & 31.00 & 23.66 & 32.88 & 32.76 \\

\midrule
\multirow{10}{*}{30}
& 1  & 23.47 & 14.10 & 22.72 & 23.62 & 17.85 & 22.79 & 23.64 \\
& 2  & 25.32 & 15.24 & 19.78 & 25.28 & 16.91 & 25.27 & 23.11 \\
& 3  & 19.87 & 16.66 & 17.87 & 22.72 & 15.63 & 19.75 & 22.74 \\
& 4  & 23.70 & 19.96 & 22.85 & 17.17 & 19.88 & 17.85 & 23.01 \\
& 5  & 22.75 & 19.66 & 22.76 & 22.86 & 22.86 & 22.75 & 23.51 \\
& 6  & 22.80 & 15.41 & 19.63 & 18.24 & 13.12 & 15.76 & 22.88 \\
& 7  & 19.91 & 16.72 & 18.17 & 19.69 & 16.63 & 22.42 & 19.85 \\
& 8  & 19.90 & 19.85 & 19.90 & 19.90 & 19.90 & 19.90 & 22.55 \\
& 9  & 22.62 & 16.57 & 18.26 & 22.74 & 14.62 & 19.81 & 22.62 \\
& 10 & 23.39 & 17.20 & 23.54 & 23.60 & 18.35 & 22.74 & 23.48 \\
\midrule
\end{tabular}}
\end{table}

\begin{table}[H]
\centering
\caption{Values of $F_A^{\min}$ and $F_A^{\text{mean}}$ for different $\lambda$ settings.}
\label{tab:lambda_scenario_A}
\resizebox{\linewidth}{!}{%
\begin{tabular}{c c | ccccc | ccccc}
\toprule
Network size & Instance
& \multicolumn{5}{c|}{$F_A^{\min}$}
& \multicolumn{5}{c}{$F_A^{\text{mean}}$} \\

\cmidrule(lr){3-7} \cmidrule(lr){8-12}

& 
& Baseline & $\lambda=0$ & $\lambda=0.5$ &  $\lambda=1.5$ & $\lambda=2$
& Baseline & $\lambda=0$ & $\lambda=0.5$  & $\lambda=1.5$ & $\lambda=2$ \\

\midrule
10 & 1  & 117.00 & 117.00 & 117.00  & 117.00 & 26.00 & 124.22 & 124.22 & 124.22  & 124.22 & 101.83 \\
   & 2  & 104.00 & 52.00  & 104.00  & 104.00 & 26.00 & 121.33 & 78.72  & 121.33  & 121.33 & 99.67 \\
   & 3  & 52.00  & 52.00  & 52.00    & 52.00  & 26.00 & 95.33  & 95.33  & 95.33    & 95.33  & 109.78 \\
   & 4  & 117.00 & 117.00 & 117.00  & 117.00 & 39.00 & 137.22 & 131.44 & 137.22  & 137.22 & 115.56 \\
   & 5  & 104.00 & 117.00 & 104.00  & 104.00 & 58.50 & 137.22 & 134.33 & 137.22  & 137.22 & 116.28 \\
   & 6  & 117.00 & 117.00 & 117.00  & 117.00 & 26.00 & 134.33 & 134.33 & 134.33  & 134.33 & 115.56 \\
   & 7  & 117.00 & 52.00  & 117.00  & 117.00 & 58.50 & 137.22 & 88.11  & 137.22   & 137.22 & 113.39 \\
   & 8  & 52.00  & 52.00  & 52.00    & 52.00  & 13.00 & 88.11  & 109.78 & 88.11  &  88.11  & 93.17 \\
   & 9  & 117.00 & 117.00 & 117.00  & 117.00 & 58.50 & 131.44 & 130.00 & 131.44  & 131.44 & 114.83 \\
   & 10 & 52.00  & 52.00  & 52.00    & 52.00  & 58.50 & 125.67 & 109.78 & 125.67 &  125.67 & 109.78 \\
\midrule

15 & 1  & 52.00  & 52.00  & 52.00    & 52.00  & 13.00 & 89.61  & 90.54  & 89.6071 &  89.61 & 99.82 \\
   & 2  & 104.00 & 104.00 & 39.00   & 58.50  & 19.50 & 132.79 & 132.79 & 79.8571 &  99.82 & 108.18 \\
   & 3  & 52.00  & 52.00  & 52.00    & 52.00  & 26.00 & 104.46 & 91.00  & 104.4640 &  104.46 & 102.61 \\
   & 4  & 52.00  & 39.00  & 52.00    & 58.50  & 13.00 & 91.93  & 72.43  & 91.9286 &  89.11 & 105.39 \\
   & 5  & 39.00  & 39.00  & 39.00    & 39.00  & 19.50 & 84.50  & 82.64  & 94.2500 &  84.50 & 102.61 \\
   & 6  & 39.00  & 58.50  & 39.00    & 52.00  & 52.00 & 84.50  & 101.68 & 84.5000 &  84.50 & 106.32 \\
   & 7  & 39.00  & 39.00  & 39.00    & 52.00  & 13.00 & 90.07  & 73.82  & 90.0714 &  90.07 & 99.82 \\
   & 8  & 52.00  & 52.00  & 52.00    & 52.00  & 19.50 & 87.75  & 84.96  & 87.7500 &  87.75 & 112.36 \\
   & 9  & 104.00 & 104.00 & 104.00  & 58.50  & 39.00 & 135.57 & 136.50 & 137.4290 &  135.57 & 111.43 \\
   & 10 & 58.50  & 52.00  & 52.00    & 58.50  & 58.50 & 97.96  & 85.43  & 87.2857 &  97.96 & 109.57 \\
\midrule

20 & 1  & 58.50 & 52.00 & 52.00  & 58.50 & 39.00 & 90.66 & 84.84 & 100.9210 &  99.89 & 111.87 \\
   & 2  & 52.00 & 58.50 & 52.00  & 52.00 & 26.00 & 98.53 & 111.18 & 98.5263 &  102.29 & 108.11 \\
   & 3  & 52.00 & 52.00 & 52.00  & 52.00 & 52.00 & 103.32 & 82.79 & 98.8684 &  103.32 & 105.03 \\
   & 4  & 58.50 & 58.50 & 52.00  & 58.50 & 13.00 & 109.47 & 99.89 & 98.1842 &  93.05 & 113.92 \\
   & 5  & 52.00 & 52.00 & 52.00  & 52.00 & 6.50  & 109.13 & 86.55 & 106.3950 &  98.87 & 109.82 \\
   & 6  & 39.00 & 58.50 & 39.00  & 58.50 & 6.50  & 84.84  & 97.84 & 84.8421 & 105.71 & 103.66 \\
   & 7  & 52.00 & 39.00 & 58.50  & 58.50 & 26.00 & 90.66  & 84.16 & 100.9210 & 95.79 & 106.05 \\
   & 8  & 52.00 & 52.00 & 58.50  & 58.50 & 39.00 & 90.32  & 95.11 & 92.7105 &  89.97 & 109.13 \\
   & 9  & 39.00 & 58.50 & 26.00  & 58.50 & 26.00 & 93.74  & 98.18 & 87.2368 &  100.58 & 112.21 \\
   & 10 & 52.00 & 58.50 & 58.50  & 52.00 & 26.00 & 96.82  & 92.03 & 97.5000 &  96.82 & 115.29 \\
\midrule

30 & 1  & 52.00 & 52.00 & 58.50  & 52.00 & 52.00 & 104.90 & 85.17 & 104.6720 &  111.40 & 103.33 \\
   & 2  & 58.50 & 52.00 & 58.50  & 58.50 & 19.50 & 109.16 & 86.52 & 109.8280 & 119.24 & 105.12 \\
   & 3  & 58.50 & 58.50 & 58.50  & 58.50 & 0.00  & 104.67 & 103.10 & 109.1550 &  109.16 & 98.62 \\
   & 4  & 58.50 & 58.50 & 58.50  & 58.50 & 6.50  & 101.76 & 94.14 & 101.7590 &111.62 & 104.90 \\
   & 5  & 52.00 & 39.00 & 52.00  & 52.00 & 39.00 & 107.59 & 91.67 & 93.6897 & 103.55 & 100.64 \\
   & 6  & 58.50 & 52.00 & 65.00  & 58.50 & 6.50  & 102.88 & 87.64 & 97.7241 & 102.88 & 104.90 \\
   & 7  & 58.50 & 58.50 & 52.00  & 58.50 & 13.00 & 114.53 & 89.43 & 109.3790 &  110.50 & 108.71 \\
   & 8  & 58.50 & 58.50 & 58.50  & 58.50 & 26.00 & 109.60 & 96.60 & 103.5520 &  113.86 & 108.26 \\
   & 9  & 58.50 & 52.00 & 58.50  & 58.50 & 6.50  & 108.71 & 89.21 & 108.7070 &  108.71 & 110.28 \\
   & 10 & 52.00 & 26.00 & 52.00  & 65.00 & 0.00  & 109.16 & 79.12 & 101.0860 &  118.79 & 110.50 \\

\bottomrule
\end{tabular}}
\end{table}

\begin{table}[H]
\centering
\caption{Values of $F_B^{\min}$ and $F_B^{\text{mean}}$ for different $\lambda$ settings.}
\label{tab:lambda_scenario_B}
\resizebox{\linewidth}{!}{%
\begin{tabular}{c c | ccccc | ccccc}
\toprule
Network size & Instance
& \multicolumn{5}{c|}{$F_B^{\min}$}
& \multicolumn{5}{c}{$F_B^{\text{mean}}$} \\

\cmidrule(lr){3-7} \cmidrule(lr){8-12}

& 
& Baseline & $\lambda=0$ & $\lambda=0.5$ &  $\lambda=1.5$ & $\lambda=2$
& Baseline & $\lambda=0$ & $\lambda=0.5$  & $\lambda=1.5$ & $\lambda=2$ \\

\midrule
10 & 1  & 26.00 & 26.00 & 26.00  & 26.00 & 8.67 & 60.43 & 60.43 & 60.43  & 60.43 & 80.41 \\
   & 2  & 26.00 & 39.00 & 26.00  & 26.00 & 8.67 & 55.35 & 61.39 & 55.35  & 55.35 & 78.00 \\
   & 3  & 32.50 & 32.50 & 32.50  & 32.50 & 8.67 & 72.46 & 72.46 & 72.46  & 72.46 & 83.30 \\
   & 4  & 29.25 & 29.25 & 29.25  & 29.25 & 13.00 & 68.83 & 63.05 & 68.83  & 68.83 & 89.56 \\
   & 5  & 31.20 & 23.40 & 31.20  & 31.20 & 26.00 & 62.45 & 66.52 & 62.45  & 62.45 & 87.75 \\
   & 6  & 23.40 & 31.20 & 23.40  & 23.40 & 13.00 & 66.52 & 64.86 & 66.52  & 66.52 & 91.48 \\
   & 7  & 31.20 & 29.25 & 31.20  & 31.20 & 19.50 & 66.30 & 55.49 & 66.30  & 66.30 & 86.91 \\
   & 8  & 29.25 & 29.25 & 29.25  & 29.25 & 13.00 & 55.48 & 47.52 & 55.49 & 55.49 & 75.83 \\
   & 9  & 26.00 & 26.00 & 26.00  & 26.00 & 13.00 & 64.28 & 62.11 & 64.28  & 64.28 & 87.39 \\
   & 10 & 29.25 & 29.25 & 29.25  & 29.25 & 19.50 & 55.11 & 47.52 & 55.11  & 55.11 & 83.42 \\
\midrule
15 & 1  & 29.25 & 29.25 & 29.25  & 29.25 & 6.50 & 62.35 & 63.28 & 62.35  & 62.35 & 85.35 \\
   & 2  & 18.57 & 18.57 & 29.25  & 29.25 & 8.67 & 47.10 & 47.38 & 51.06  & 71.89 & 84.42 \\
   & 3  & 29.25 & 29.25 & 29.25  & 29.25 & 8.67 & 68.39 & 60.50 & 68.39  & 68.39 & 87.21 \\
   & 4  & 26.00 & 29.25 & 26.00  & 26.00 & 6.50 & 62.45 & 52.70 & 62.45  & 60.59 & 79.74 \\
   & 5  & 29.25 & 29.25 & 29.25 & 29.25 & 13.00 & 62.12 & 60.26 & 58.33  & 62.12 & 81.64 \\
   & 6  & 39.00 & 26.00 & 39.00 & 39.00 & 26.00 & 67.79 & 61.19 & 67.79  & 67.79 & 89.38 \\
   & 7  & 32.50 & 32.50 & 32.50  & 32.50 & 4.33 & 65.62 & 59.12 & 65.62  & 65.62 & 86.43 \\
   & 8  & 29.25 & 29.25 & 29.25  & 29.25 & 9.75 & 65.14 & 55.25 & 65.14  & 65.14 & 84.50 \\
   & 9  & 19.50 & 21.67 & 17.33 & 19.50 & 13.00 & 49.41 & 49.12 & 49.26  & 49.41 & 86.28 \\
   & 10 & 26.00 & 32.50 & 32.50 &  26.00 & 19.50 & 57.01 & 66.08 & 67.94  & 57.01 & 81.02 \\
\midrule
20 & 1  & 26.00 & 26.00 & 26.00 &  29.25 & 3.90 & 69.96 & 54.84 & 72.18 & 67.16 & 85.56 \\
   & 2  & 26.00 & 21.67 & 26.00 &  26.00 & 8.67 & 69.45 & 71.86 & 69.45 & 76.80 & 88.72 \\
   & 3  & 26.00 & 26.00 & 26.00 &  23.40 & 17.33 & 66.43 & 53.30 & 63.73  & 72.56 & 85.75 \\
   & 4  & 26.00 & 26.00 & 26.00 &  26.00 & 4.33 & 77.32 & 64.42 & 63.29 &  69.45 & 86.95 \\
   & 5  & 26.00 & 19.50 & 26.00 &  26.00 & 2.17 & 78.57 & 56.37 & 75.61 &  76.12 & 85.01 \\
   & 6  & 29.25 & 26.00 & 29.25 &  26.00 & 2.17 & 66.20 & 63.46 & 66.20 &  78.86 & 87.46 \\
   & 7  & 26.00 & 29.25 & 19.50 &  26.00 & 8.67 & 66.20 & 69.79 & 60.53 &  68.08 & 88.55 \\
   & 8  & 26.00 & 26.00 & 29.25 &  29.25 & 13.00 & 70.30 & 64.49 & 78.51 & 75.95 & 88.89 \\
   & 9  & 26.00 & 21.67 & 26.00 &  21.67 & 8.67 & 64.66 & 64.71 & 63.12 &  76.63 & 85.98 \\
   & 10 & 29.25 & 29.25 & 29.25 &  29.25 & 8.67 & 74.68 & 70.23 & 75.71 &  75.02 & 87.46 \\
\midrule
30 & 1  & 19.50 & 17.33 & 16.25 &  16.25 & 8.36 & 65.69 & 53.87 & 66.55 &  71.37 & 84.49 \\
   & 2  & 21.67 & 16.25 & 19.50 & 19.50 & 8.67 & 78.67 & 55.18 & 82.67 &  78.59 & 88.12 \\
   & 3  & 15.60 & 19.50 & 14.63 &  14.63 & 0.00 & 67.78 & 69.93 & 67.67  & 67.67 & 85.28 \\
   & 4  & 19.50 & 19.50 & 19.50 &  19.50 & 0.65 & 75.03 & 64.66 & 75.03  & 80.76 & 83.25 \\
   & 5  & 19.50 & 19.50 & 21.67 &  19.50 & 2.17 & 68.89 & 53.79 & 66.34  & 70.45 & 79.97 \\
   & 6  & 19.50 & 14.63 & 21.67 & 16.25 & 2.17 & 71.12 & 56.92 & 67.75  & 64.22 & 85.28 \\
   & 7  & 15.60 & 12.00 & 13.00  & 13.00 & 2.79 & 68.88 & 55.63 & 60.99  & 63.50 & 83.49 \\
   & 8  & 16.25 & 16.25 & 19.50  & 19.50 & 0.43 & 67.86 & 58.61 & 61.10  & 72.40 & 85.38 \\
   & 9  & 18.57 & 19.50 & 18.57  & 18.57 & 0.00 & 63.74 & 53.96 & 63.74  & 63.74 & 83.51 \\
   & 10 & 19.50 & 16.71 & 19.50  & 19.50 & 0.00 & 62.08 & 50.60 & 66.00  & 62.21 & 85.78 \\

\bottomrule
\end{tabular}}
\end{table}

\begin{table}[H]
\centering
\caption{Values of $F_C^{\min}$ and $F_C^{\text{mean}}$ for different $\lambda$ settings.}
\label{tab:lambda_scenario_C}
\resizebox{\linewidth}{!}{%
\begin{tabular}{c c | ccccc | ccccc}
\toprule
Network size & Instance
& \multicolumn{5}{c|}{$F_C^{\min}$}
& \multicolumn{5}{c}{$F_C^{\text{mean}}$} \\

\cmidrule(lr){3-7} \cmidrule(lr){8-12}

& 
& Baseline & $\lambda=0$ & $\lambda=0.5$ &  $\lambda=1.5$ & $\lambda=2$
& Baseline & $\lambda=0$ & $\lambda=0.5$  & $\lambda=1.5$ & $\lambda=2$ \\

\midrule
10 & 1  & 43.88 & 43.88 & 43.88 &   43.88 & 11.14 & 7.82 & 7.82 & 7.82 &  7.82 & 9.24 \\
   & 2  & 43.88 & 43.88 & 43.88 &   43.88 & 11.14 & 7.46 & 7.06 & 7.46  & 7.46 & 8.98 \\
   & 3  & 36.56 & 36.56 & 36.56 &  36.56 & 11.14 & 8.38 & 8.38 & 8.38 &  8.38 & 9.64 \\
   & 4  & 43.88 & 43.88 & 43.88 & 43.88 & 16.71 & 8.85 & 8.21 & 8.85 &  8.85 & 10.32 \\
   & 5  & 55.71 & 55.71 & 55.71 &  55.71 & 32.91 & 8.26 & 8.07 & 8.26 &  8.26 & 10.16 \\
   & 6  & 42.12 & 42.12 & 42.12 &  42.12 & 14.62 & 8.56 & 8.56 & 8.56  & 8.56 & 10.51 \\
   & 7  & 48.75 & 48.75 & 48.75 & 48.75 & 25.07 & 8.61 & 8.43 & 8.61 & 8.61 & 10.04 \\
   & 8  & 43.88 & 43.88 & 43.88 & 43.88 & 13.00 & 6.68 & 6.68 & 6.68 &  6.68 & 8.68 \\
   & 9  & 46.80 & 46.80 & 46.80 &  46.80 & 25.07 & 8.32 & 8.23 & 8.32 &  8.32 & 10.10 \\
   & 10 & 43.88 & 43.88 & 43.88 &  43.88 & 25.07 & 7.35 & 6.39 & 7.35 &  7.35 & 9.64 \\
\midrule
15 & 1  & 35.00 & 35.00 & 35.00  & 35.00 & 7.00 & 4.62 & 4.69 & 4.62 &  4.62 & 6.18 \\
   & 2  & 31.50 & 32.50 & 35.00  & 31.50 & 10.11 & 3.98 & 4.00 & 3.83 & 5.30 & 6.16 \\
   & 3  & 31.50 & 31.50 & 31.50  & 31.50 & 10.11 & 5.11 & 4.51 &  5.11 & 5.11 & 6.31 \\
   & 4  & 31.50 & 31.50 & 31.50  & 31.50 & 7.00 & 4.64 & 3.88 & 4.64 &  4.51 & 5.84 \\
   & 5  & 31.50 & 31.50 & 31.50  & 31.50 & 14.00 & 4.57 & 4.44 & 4.39 & 4.57 & 5.95 \\
   & 6  & 39.00 & 40.44 & 39.00 &  39.00 & 28.00 & 4.94 & 4.64 & 4.94 & 4.94 & 6.48 \\
   & 7  & 35.00 & 35.00 & 35.00 & 35.00 & 5.06 & 4.83 & 4.31 & 4.83 & 4.83 & 6.25 \\
   & 8  & 31.50 & 31.50 & 31.50  & 31.50 & 10.50 & 4.79 & 4.12 & 4.79 &  4.79 & 6.19 \\
   & 9  & 36.40 & 32.76 & 36.40 & 36.40 & 15.17 & 4.17 & 4.14 & 4.21 &  4.17 & 6.30 \\
   & 10 & 40.44 & 35.00 & 35.00  & 40.44 & 22.75 & 4.35 & 4.83 & 4.96 &  4.35 & 5.95 \\
\midrule
20 & 1  & 30.88 & 27.44 & 27.44  & 30.88 & 7.41 & 3.75 & 2.98 & 3.89 &  3.64 & 4.59 \\
   & 2  & 29.06 & 29.06 & 29.06  & 29.06 & 9.69 & 3.74 & 3.91 & 3.74 &  4.12 & 4.73 \\
   & 3  & 29.06 & 27.44 & 29.06  & 29.06 & 19.37 & 3.61 & 2.90 & 3.46 &  3.91 & 4.57 \\
   & 4  & 29.06 & 29.06 & 29.06 &  29.06 & 4.84 & 4.17 & 3.51 & 3.44 &  3.73 & 4.66 \\
   & 5  & 27.44 & 24.22 & 29.06 &  27.44 & 2.42 & 4.23 & 3.06 & 4.07 & 4.08 & 4.55 \\
   & 6  & 29.06 & 29.06 & 30.88 &  29.06 & 2.42 & 3.54 & 3.45 & 3.54 & 4.23 & 4.65 \\
   & 7  & 27.44 & 30.88 & 24.22 &  27.44 & 9.69 & 3.57 & 3.72 & 3.33 &  3.67 & 4.71 \\
   & 8  & 30.88 & 29.06 & 30.88  & 30.88 & 14.53 & 3.76 & 3.49 & 4.17 &  4.04 & 4.74 \\
   & 9  & 24.22 & 24.22 & 26.00  & 24.22 & 9.69 & 3.49 & 3.51 & 3.40 & 4.10 & 4.60 \\
   & 10 & 30.88 & 30.88 & 30.88  & 30.88 & 9.69 & 4.00 & 3.76 & 4.05 &  4.02 & 4.69 \\
\midrule
30 & 1  & 21.75 & 21.75 & 18.10  & 18.13 & 10.54 & 2.32 & 1.90 & 2.38 &  2.51 & 2.94 \\
   & 2  & 23.27 & 18.13 & 20.94  & 20.94 & 10.10 & 2.75 & 1.94 & 2.88 &  2.76 & 3.06 \\
   & 3  & 18.10 & 21.75 & 16.31  & 16.31 & 0.00 & 2.38 & 2.45 & 2.39 &  2.39 & 2.96 \\
   & 4  & 21.75 & 20.94 & 21.75 & 20.94 & 0.94 & 2.62 & 2.27 & 2.62 &  2.82 & 2.90 \\
   & 5  & 20.94 & 20.94 & 23.27 & 21.75 & 5.12 & 2.43 & 1.90 & 2.32 & 2.47 & 2.80 \\
   & 6  & 20.94 & 16.31 & 23.27  & 20.94 & 2.33 & 2.49 & 2.00 & 2.37 &  2.26 & 2.97 \\
   & 7  & 18.10 & 15.71 & 15.71  & 18.10 & 5.24 & 2.43 & 1.97 & 2.17 &  2.25 & 2.91 \\
   & 8  & 18.13 & 18.13 & 20.94 & 20.94 & 5.05 & 2.39 & 2.07 & 2.16 &  2.55 & 2.97 \\
   & 9  & 20.94 & 20.94 & 20.94  & 20.94 & 0.84 & 2.26 & 1.91 & 2.26 &  2.26 & 2.91 \\
   & 10 & 21.75 & 20.20 & 21.75  & 21.75 & 0.00 & 2.20 & 1.78 & 2.32 & 2.22 & 2.99 \\

\bottomrule
\end{tabular}}
\end{table}

\end{document}